\documentclass[11pt]{article}
\usepackage{geometry}                
\usepackage{makeidx}
\usepackage[pdftex]{graphicx}
\usepackage{lscape}
\usepackage{amssymb}
\usepackage{epstopdf}
\usepackage{amsmath,amssymb,amsthm}
\usepackage{latexsym}
\usepackage{listings}
\usepackage{chemarr}
\pdfoutput=1

\newtheorem{axiom}{Axiom}[section]
\newtheorem{defin}{Definition}[section]
\newtheorem{theo}{Theorem}[section]
\newtheorem{propo}{Proposition}[section]
\newtheorem{lemma}{Lemma}[section]
\newtheorem{corol}{Corollary}[section]
\newtheorem{remark}{Remark}[section]
\newcommand{\bax}{\begin{axiom}}
\newcommand{\eax}{\end{axiom}}
\newcommand{\bass}{\begin{assumption}}
\newcommand{\eass}{\end{assumption}}
\newcommand{\bdefi}{\begin{defin}}
\newcommand{\edefi}{\end{defin}}
\newcommand{\bth}{\begin{theo}}
\renewcommand{\eth}{\end{theo}}
\newcommand{\bprop}{\begin{propo}}
\newcommand{\eprop}{\end{propo}}
\newcommand{\blem}{\begin{lemma}}
\newcommand{\elem}{\end{lemma}}
\newcommand{\bcor}{\begin{corol}}
\newcommand{\ecor}{\end{corol}}
\newcommand{\brem}{\begin{remark}}
\newcommand{\erem}{\end{remark}}
\newcommand{\bpf}{\begin{proof}}
\newcommand{\epf}{\end{proof}}

\newcommand{\noi}{\noindent}
\newcommand{\bn}{{\bf n}}

\newcommand{\bk}{{\bf k}}
\newcommand{\bx}{{\bf x}}
\newcommand{\vde}{{\vec{\de}}}
\newcommand{\field}[1]{\mathbb{#1}}

\newcommand{\R}{\field{R}}

\newcommand{\Z}{\field{Z}}
\newcommand{\N}{\field{N}}
\newcommand{\X}{\field{X}}
\newcommand{\LL}{\field{L}}
\newcommand{\pa}{\partial}
\newcommand{\tr}{\mbox{tr }}
\newcommand{\ra}{\rightarrow}

\newcommand{\de}{\delta}

\newcommand{\sig}{\sigma}
\newcommand{\Sig}{\Sigma}
\newcommand{\eps}{\epsilon}
\newcommand{\veps}{\varepsilon}
\newcommand{\A}{{\mathcal A}}
\newcommand{\I}{{\mathcal I}}
\newcommand{\Y}{{\mathcal Y}}

\newcommand{\Eop}{{\bf E}}
\newcommand{\Dop}{{\bf \Delta}}
\newcommand{\Lop}{{\widehat{{\mathcal L}}}}
\newcommand{\Aop}{{\widehat{{\mathcal A}}}}
\newcommand{\Pro}{{{{\mathcal P}}}}
\newcommand{\Kop}{{{\mathcal K}}}
\newcommand{\Ld}{{{{\mathcal L}}}}
\newcommand{\Kd}{{{{\mathcal K}}}}

\newcommand{\id}{{\bf id}}
\newcommand{\bPi}{{\bf \Pi}}
\newcommand{\bpi}{{\bf \pi}}
\newcommand{\be}{{\bf 1}}
\newcommand{\bI}{{\bf I}}
\newcommand{\idf}{\emph{i.d.f.}}
\newcommand{\fdf}{\emph{f.d.f.}}
\newcommand{\app}{\approx}
\newcommand{\lb}{\langle}
\newcommand{\rb}{\rangle}

\begin{document}

\title{Molecular Systems with Infinite and Finite Degrees of Freedom. Part I: Multi-Scale Analysis}


\author{
Luca Sbano, 
Mathematics Institute, University of Warwick \\
              \texttt{sbano@maths.warwick.ac.uk}\\         
              and\\     
              Markus Kirkilionis
              Mathematics Institute, University of Warwick \\
              \texttt{mak@maths.warwick.ac.uk}  
}

\date{}

\maketitle

\begin{abstract}
The paper analyses stochastic systems  describing reacting molecular systems with a combination of two types of state spaces, a finite-dimensional, and an infinite dimenional part. As a typical situation  consider the interaction of larger macro-molecules, finite and small in numbers per cell  (like protein complexes), with smaller, very abundant molecules, for example metabolites. We study the construction of the continuum approximation of the associated Master Equation (ME) by using the Trotter approximation \cite{Trotter}. The continuum limit shows regimes where the finite degrees of freedom evolve faster 
than the infinite ones. Then we develop a rigourous asymptotic adiabatic theory upon the condition that the jump process arising from the finite degrees of freedom of the Markov Chain (MC, typically describing conformational changes of the macro-molecules) occurs with large frequency. In a second part of this work, the theory is applied to derive typical enzyme kinetics in an alternative way and interpretation within this framework.
\end{abstract}

\section{Introduction}

Think of a typical situation in Cell Biology, the interaction of macro-molecules in the cell. In most cases there will be a small number of macro-molecular machines, like enzymes, ion-channels, polymerases, ribosomes etc. which are essential for cellular function, but which will not be very abundant in numbers per cell. Moreover this number will typically not change over time of observation. These machines will have different states of operation, like an ion channel can be closed or open. The states of operation of such machines can in general be described by finitely many different discrete states. These discrete states can be associated with meta-stable conformations of proteins (see for example \cite{deuflhard}). Smaller molecules like ions, or metabolites like sugars, will interact with these macro-molecules. The most typical and best studied situation are enzymes catalysing metabolic reactions. The classical way to describe the resulting kinetics is given in \cite{siegel}, see also \cite{keener}. The number of these smaller molecules clustered in different species will change typically over time of observation. Assuming no inherent spatial structure of the process, this gives rise to coupled systems of two well studied mathematical objects, Markov chains describing the transitions between the different modes of operation of the macro-molecules, and birth-death processes with reactions describing the change in numbers of the smaller abundant molecules. In this paper we will study both mathematical objects simultaneously as one system, giving a rigourous derivation of the continuum limit. With other interpretations the theory can also applied in various other fields of sciences where interaction of different finite state 'machines' will occur, like epidemiology, manufacturing or  economy.

\begin{figure}[htbp] 
   \centering
   \includegraphics[width=7cm]{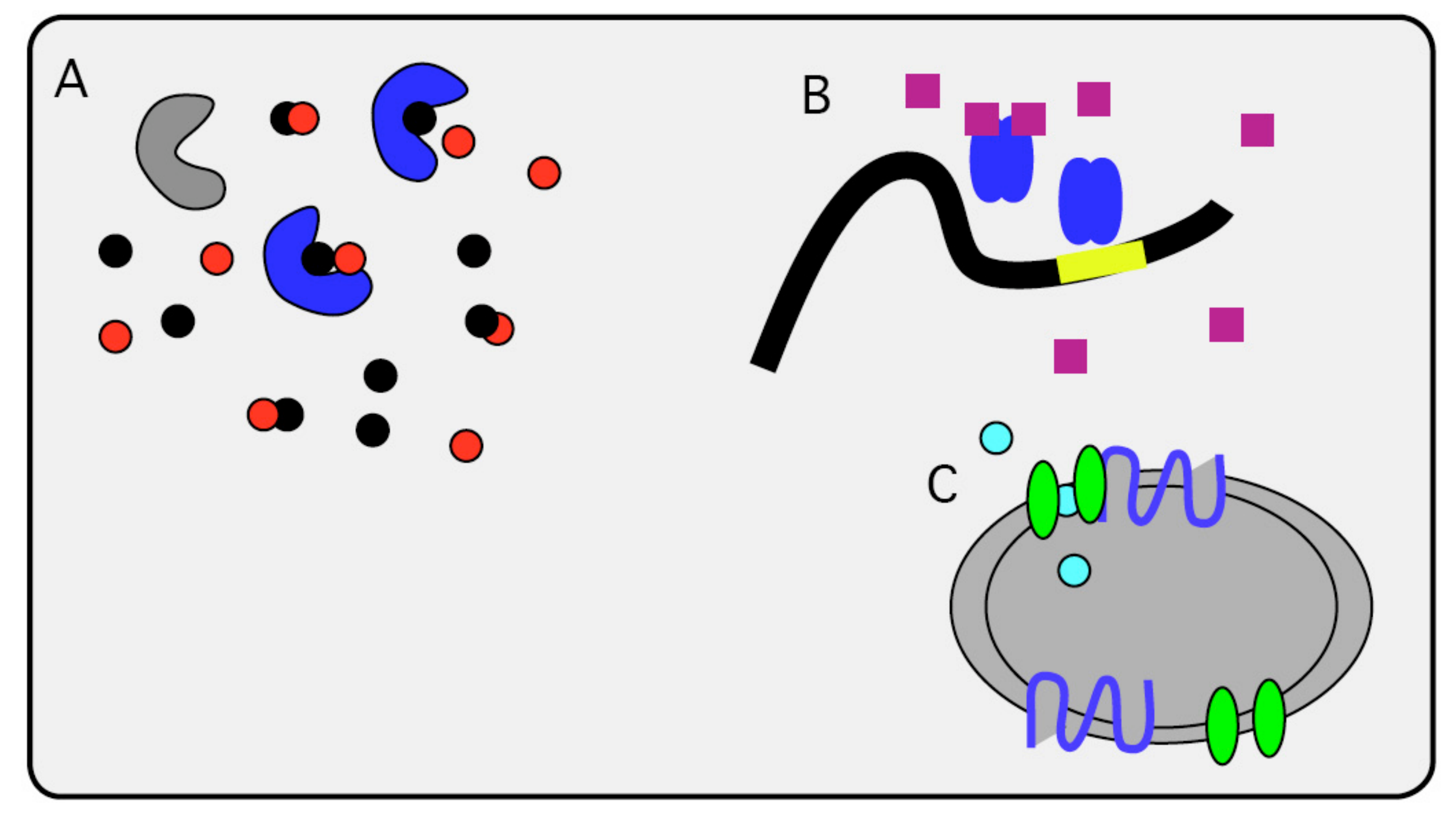} 
   \caption{\small Different typical interactions of small and large molecules in a typical cell. (A): Interaction of enzymes with metabolites. The product is a molecule consisting of two elementary species. Enzymes react as catalysators. (B): Genetic interactions. For example a repressor can bind to the DNA only in the case it is in a conformation characterised by the absence of smaller 'inducing' molecules. (C): Membrane proteins act in various ways as molecular machines, for example as ion channels.}
   \label{fig:example}
\end{figure}

As a concrete example from Genetics assume mRNA is transcribed depending on whether a specific DNA binding site is bound to a transcription factor $A$. There are two possible \emph{limit} regimes: either no $A$ molecules are  binding causing no mRNA transcription, or molecules $A$ are binding to the DNA implying a maximum transcription rate for the mRNA. Usually such binding/unbinding events occur at large frequencies, and they are proportional to the concentration of the transcription factor. This leads to an effective transcription rate resulting from an  "effective average" of the binding/unbuinding event depending on the concentration of $A$. This effect is usually modelled by a Hill-type kinetics (\cite{CCB}): 
  
  \[K([A])=\frac{K_1\,[A]^n}{1+K_2\,[A]^n}\]
  
This kinetics describes an effective reaction rate with saturation behaviour for large concentrations $[A]$  of the transcription factor $A$. Here $n$ is a positive integer, an exponent controlling  the slope of the sigmoidal $K([A])$. In this paper we consider as an illustrative example the case $n=1$, in \cite{sbano2} we shall describe the general case. We will follow this particular example throughout this paper (part I) starting from microscopic assumptions, and derive the above deterministic limit in part II.

\subsubsection*{Birth-death processes and the density assumption}

We will assume that at time $t$ the state of the subsystem describing small molecules of variable number is fully determined by the collection of these numbers belonging to different species. The time evolution is given through the transition from one possible collection of these molecules to another. The transition itself is prescribed by the reactions, which contain the rates at which the various species form complexes, i.e. other species. Reactions define the dynamics by providing the probability rates at which the elementary transitions occur, and this leads  to constructing a stochastic process with independent increments (Markov property).  For this part of the system the state at time $t$ is described by a probability distribution,  which solves the so-called Master Equation (ME) (see for example \cite{Vankampen} and \cite{gardiner}). The ME is a differential equation with respect to time and a difference equation with respect to the various particle numbers. In many situations the number of molecules involved is of the order of the  Avogadro number $\sim 10^{23}$, which implies that the concept of density can be introduced.  That can be achieved by constructing the so-called  continuum approximation of the ME, an operation that transforms the ME into a Fokker-Planck Equation (FPE). This correspondence is a very delicate point and often neglected in applications. Usually the continuum approximation is investigated through the \emph{Kramers-Moyal} expansion or the \emph{Van Kampen} size-expansion \cite{gardiner}, \cite{Vankampen}.  Here we employ the Trotter approximation \cite{Trotter,Kurtz,Pazy} to study multi-scale reactions systems. In fact the stochastic process generated by the FPE is a valid approximation to the one generated by the ME only for a fixed time interval (for further details see \cite{Kurtz1}, \cite{Kurtz2},\cite{Kurtz3}, and \cite{Hanggi} for a recent review).  A representation of the ME can be directly studied numerically through Gillespie's algorithm (see \cite{Gillespie1}) which provides the construction of the underlying stochastic process. Our focus as outlined in the reminder of this introduction aims at understanding  how  different scales are affecting the macroscopic dynamics. These are the ingredients of models usually of interest in Mathematical Biology or Systems Biology.

\subsubsection*{Molecular subsystems with finite degrees of freedom, mixed systems}

To summarise the system to study will have two kind of state variables (degrees of freedom), first variables which admit a \emph{continuum approximation} as just described, and secondly variables whose discrete nature is essential. Earlier work to study such systems include \cite{kepler-elston} and \cite{sasai-wolynes}.  In this class the state is described by a set of numbers of molecules, and a set of  finite discrete states describing all possible molecular conformations, binding/unbinding events, etc. The dynamics is described by a master equation and its solution, a probability distribution generating a Markov process.  After taking the continuum approximation it has been formally shown in an appendix of \cite{kepler-elston} that  the whole process is a combination of a continuous process and a discrete Markov chain with finite states. Such systems are sometimes also called \emph{random evolutions} \cite{hersh}. They are ubiquitous in the realm of complex systems. It may happen that infinite and finite degrees of freedom evolve on different time scales. Indeed in many applications like \cite{kepler-elston} it appears that the dynamics associated to the finite degrees of freedom evolves very fast, and therefore it is important to understand how this affects the dynamics associated to the infinite degrees of freedom. Heuristically this is usually done by introducing  \emph{ad hoc} reaction rates (see \cite{CCB}), which mimic some sort of "averaged effect". \\
 
Some remarks on this previous work. In \cite{kepler-elston} such systems are studied by using the ME method and by taking a heuristic continuum approximation. Subsequently  formal asymptotic methods are applied to study the large frequency problem. 
In the context of simulation the analysis of systems with fast and slow dynamics has been addressed in \cite{Petzold}, where  the updates of  the infinite state variables was optimised. Collective effects of many particles in biological systems have also been investigated  in \cite{sasai-wolynes} where a many-particle method is used. This approach has its origins in multi-body and field theory, (see \cite{doi1}). Inspired by the approach presented in an appendix of \cite{kepler-elston}  we develop in this paper a rigourous analysis  of the adiabatic theory for systems with infinite and finite degrees of freedoms. 
The paper presents a general formulation for the ME for such systems whose state space is formed by $N$ types of particles and $g$ finite states. Let $\Sigma$ be the set of all possible discrete states. At time $t$ the system is in a state 
$(\bn,\sigma)\in \N^N \times \Sig$ with probability $P_\sig(\bn,t)$.  The $N$-tuple $\bn=(n_1,...,n_N)\in\N^N$ represents the collection of free molecules of different species, and $\sigma$ is a discrete state in $\Sigma$ describing the conformational changes of macro-molecules which number does not change during system observation. The ME for the probability  $P_s(\bn,t)$ is then given by

\begin{equation}
\frac{\pa P_\sig(\bn,t)}{\pa t}=\sum_{\sig'\in\Sig}\Ld^*_{\sig\sig'}(\bn)(P_{\sig'}(\bn,t))+\sum_{\sig'\in \Sig}\Kd^T_{\sig\sig'}(\bn)\,P_{\sig'}(\bn,t)\mbox{ with $\sig\in \Sig$,}
\label{main-ME00}
\end{equation}

where $\Ld^*(\bn)$ is a collection of difference operators  (the '$^*$' indicating the adjoint of an operator, a notation which will become clear later in the paper) and $\Kd^T(\bn)$ is the transpose of a generator of a Markov chain on $S$. 
We will study the ME with two methods, the {\it continuum limit} and the {\it adiabatic approximation}.

\subsubsection*{The continuum limit}
The continuum limit will be formulated by using  the so-called \emph{Trotter approximation} (see \cite{Trotter}, \cite{Pazy}). Trotter's method has been also used in \cite{Kurtz,Kurtz2,Kurtz3}. The equation (\ref{main-ME00}) is derived on the base of the elementary molecular processes that depend on the scale at which they are studied. It is therefore important 
to reformulate the ME taking into account its dependence on the size and time scales. Let us define two sets of scales:

\begin{enumerate}
\item The size scales $\vde=(\de_1,...,\de_N)$, $\de_i>0$,
\item the  time scale $\tau>0$.
\end{enumerate}

Let us define the following lattice

\begin{equation}
\LL_{\vde}=\{(n_1\de_1,...,n_N\de_N):~~(n_1,...,n_N)\in\N^N\}
\end{equation}

The state of the system is now specified on $\LL_\vde\times\Sigma$. The processes on $\LL_\vde$ will be \emph{birth-death} interaction of the form

 \[(n_1\de_1,...n_i\de_i,...,n_N\de_N)\ra(n_1\de_1,...,n_i\de_i\pm\de_i,...,n_N\de_N),\]
 
 the processes on $\Sig$ are transitions
  
 \[\sig\ra\sig, '\]
 
 driven by a finite Markov chain. Note that for fixed $\vde$ a function $f$ defined on $\LL_\vde$ is 
 fully determined by its values on $n$, i.e. when $\vde$ is fixed we can consider 
 $f$ defined over $\N^N$, writing $f(\bn\,\vde)=f(\bn)$. Equation  (\ref{main-ME00})  is now rewritten in a form where $\Ld^*$ and $\Kd^*$ are  operators depending on $\vde$ and $\tau$:
 
\begin{equation}
\frac{\pa P_\sig(\bn,t)}{\pa t}=\sum_{\sig'\in\Sig}\Ld^*_{\sig\sig'}[\vde,\tau;\bn](P_{\sig'}(\bn,t))+\sum_{\sig'\in \Sig}\Kd^T_{\sig\sig'}[\vde,\tau;\bn]\,P_{\sig'}(\bn,t).
\label{main}
\end{equation}

 The continuum limit is the study of the form of  $\Ld^*$ and $\Kd^*$ as $\vde\ra 0$ and $\tau\ra 0$. Equation (\ref{main}) is defined on the space of probability measures on $\LL_\vde\times\Sigma$:
 
\[\X^*_\vde\doteq\left\{P \;  | \sum_{\bn\in\LL_\vde,\sig\in \Sig}P_\sig(\bn)=1\right\}.\]

The continuum limit is naturally constructed on the dual of $\X^*_\vde$ (see \cite{Trotter}), namely on

\[\X_\vde\doteq\left\{u:\LL_\vde\times\Sigma\ra\R^{|\Sig|} \; | \;
\sup_{\bn\in\LL_\vde,\sig\in \Sig}|u_\sig(\bn)|<\infty\right\}.\]

The pairing between $\X_\vde^*$ and $\X_\vde$ is

\[\lb P,u\rb\doteq\sum_{\bn\in\LL_\vde,\sig\in \Sig}P_\sig(\bn)u_\sig(\bn).\]

Using the duality,  the ME defined on $\X_\vde$ becomes the Kolmogorov equation

\begin{equation}
\frac{\pa u_\sig(\bn,t)}{\pa t}=
\sum_{\sig'\in\Sig}\Ld_{\sig\sig'}[\vde,\tau;\bn](u_{\sig'}(\bn,t))+\sum_{\sig'\in \Sig}\Kd^T_{\sig\sig'}[\vde,\tau;\bn]\,u_{\sig'}(\bn,t),
\label{kolmogorov}
\end{equation}

where

\[\lb \Ld^*P,u\rb=\lb P,\Ld u\rb,~~\lb \Kd^TP,u\rb=\lb P,\Kd u\rb.\]

For any fixed $\vde$ and $\tau$ equation (\ref{kolmogorov}) is an infinitesimal generator of a Markov process on $\LL_\vde\times\Sig$. Now let $\vde_n$ and $\tau_n$ two sequences of scales such that 

\[\vde_n\ra 0,~~\tau_n\ra 0\]

as $n\ra \infty$. Then we can define a sequence of spaces $\X_n=\X_{\vde_n}$, $\X^*_n=\X^*_{\vde_n}$, and a sequence of operators $\Ld_n,\Kd_n$ acting on $\X_n$. In \cite{Trotter}, a method is introduced to study the limit of $\Ld_n,\Kd_n$  as $n\ra\infty$. The idea is to look at  $\Ld_n,\Kd_n$ defined on $\X_n$ as an approximation of infinitesimal generators $\Lop,\Kd$ defined on a suitable Banach space that, in most applications, is given by $\X=C^0(\R^N,\R^g)$. Following \cite{Trotter} we construct a sequence of projections

 \[\Pro_n:\X\ra\X_n,\]
 
 and we state that a sequence $u_n\in\X_n$ approximates $u\in\X$ if
 
 \[\|\Pro_n(u)-u_n\|_n\ra 0.\]
 
This will be denoted by $u_n\app u$.  In order to take into account the presence of the scales $\vde_n$ and $\tau_n$, we modify the criteria to approximate  the limit of operators (in an adapted way different from \cite{Trotter,Pazy}) . In fact a sequence of linear operators $\A_n:\X_n\mapsto\X_n$  is now a function of  $\vde_n$ and $\tau_n$, and in general we cannot expect some  limit  to exist for any choice of $\vde_n\ra 0,\tau_n\ra 0$. For this reason we say that the sequence of operator converges, $\A_n\app \Aop$,   if there exists $\vde_n\ra 0,\tau_n\ra 0$, and $\Aop$ defined on $\X$  such that
 
 \[\|\A_n(\Pro_n(u))-\Pro_n(\Aop(u))\|_n\ra 0\mbox{ for all $u$ in the domain of $\Aop$}, \]
 
 as $n\ra \infty$. It is important to note that the limit is now depending on the choice of the  sequence of scales converging to $0$. We shall show that in general operators may have pre-factors which characterise their limit behaviour. The typical case will be  $\Ld_n\app\Lop$ and $\Kd_n\app\frac{1}{\eps}\Kd$, which yields the limit
 
 \begin{equation}
 \Aop=\Lop+\frac{1}{\eps}\Kd.
 \label{limit-op}
 \end{equation}
 
Here the constant $\eps$ will in general be a function of  $\vde_n\ra 0,\tau_n\ra 0$, so depending on the scales. This approach determines the operator (\ref{limit-op}) that, for fixed $\eps$,  is the infinitesimal generator of a process on $\R^N\times\Sig$. In many applications it turns out that $\eps$ is an infinitesimal function in 
$\vde_n\ra 0,\tau_n\ra 0$ and this leads very naturally to the study of the adiabatic approximation for the equation

\begin{equation}
\frac{\pa u(\bx,t)}{\pa t}=\Aop(u(\bx,t)).
\end{equation}

In the applications is often considered the adjoint equation, the Fokker Planck equation (FPE) which corresponds directly to the ME. The FPE is defined on the dual of $\X^*$ and reads

\begin{equation}
\frac{\pa \rho(\bx,t)}{\pa t}=\Lop^*(\bx)(\rho(\bx,t))+\frac{1}{\eps}\Kd^T(\bx)\,\rho(\bx,t).
\label{main-ME0}
\end{equation}
and
\[\X^*\doteq\left\{\rho:\sum_{\sig\in\Sig}\int d\bx\rho(\bx,\sig)u(\bx,\sig)<\infty\right\}.\]



\subsubsection*{Adiabatic approximation}

The adiabatic approximation theory is based on the observation that for sufficiently small $\eps$ the dynamics of the finite state Markov chain is faster than the one of the birth-death process. This should be a reasonable assumption for most or all macro-molecular behaviour in a cell. Such an assumption  implies that on sufficiently large time scales the Markov chain can be considered at equilibrium. Generalising \cite{kepler-elston} we assume that the Markov chain has possibly more than one stationary measure:

\[M_\Kd\doteq\{\mu(\bx)\in\X^*_\vde:\Kd^T(\bx)\mu(\bx)=0\}.\]

\noi To avoid trivialities we assume that

\[m_\Kd\doteq\dim(M_\Kd)<g.\]

\noi Any convex combination

\[\mu=\sum_{m=1}^{m_\Kd}\theta_m\,\mu^{(m)}\mbox{ with } \sum_{m=1}^{m_\Kd}\theta_m=1\mbox{ where } \theta_m\in\R_+\]

\noi is in $M_\Kd$ (see \cite{Brzezniak}). Each such measure describes the possible asymptotic behaviour of the Markov chain which is in general decomposable, i.e. a product of $m_\Kd$ Markov chains. We now take one convex combination $\mu\in M_\Kd$ and construct the adiabatic theory for the FPE obtaining an asymptotic expansion in $\eps$ of $\rho$. This expansion has a leading order term, which will be called \emph{average dynamics}. This dynamics is given by

\begin{equation}
\frac{\pa f(\bx,t)}{\pa t}=\sum_{m=1}^{m_\Kd}\sum_{\sig\in \Sig}\theta_m\Lop_\sig^*(\bx)(\mu^{(m)}_\sig(\bx)f(\bx,t)),
\label{average-dyn1}
\end{equation}

where 

\begin{equation}
f(\bx,t)=\sum_{\sig\in \Sig_\mu}\rho_\sig(\bx,t)\mbox{ with } \Sig_\mu=\{\sig\in \Sig:\mu_{\sig}\neq 0\}
\end{equation}

is the marginal distribution associated to $\mu$. It is noteworthy that the appearance of an averaged dynamics occurs in the modified Gillespie's algorithm  as presented in \cite{Petzold}.

\subsubsection*{Description in terms of ODEs and SDEs}

In the study of (\ref{main-ME0}) for small $\veps$ one could make the non-trivial observation that equation (\ref{main-ME0}) up to order $O(\veps)$ generates a Markov process  described  by a stochastic differential equation. This observation  which entails to show that  (\ref{main-ME0}) up to $O(\veps)$ reduces to a parabolic operator will be clarified in another paper.  It is important to mention here that such  an approximation is valid only on finite time interval as it was shown in  \cite{Kurtz1}, \cite{Kurtz2} and \cite{Kurtz3}. Under this restricted condition the dynamics can be described by the following Ito stochastic differential equation

\begin{equation}
dx_\alpha(t)=A_\alpha(\bx(t))\,dt+\sqrt{\veps}\sum^N_{\beta=1}\sigma_{\alpha\beta}(\bx(t))\,dw^\beta_t\mbox{ with $\alpha=1,...,N$,}
\end{equation}

where $\{w_t^\beta\}_{\beta=1}^N$ are $N$-independent Wiener processes.
 Here $\|\sigma(\veps,\bx)\|\sim\sqrt{\veps}$ and $A(\bx)$ is the \emph{averaged vector field} given by

\begin{equation}
A_\alpha(\bx)=\sum_{j\in S}\theta_m\,L^{j}_\alpha(\bx)\mu^{(m)}_j(\bx).
\label{average}
\end{equation}

Here $L^j(\bx)$ is the deterministic vector field associated to the finite state $j$. Moreover $A(x)$ is the average over the stationary measure $\mu(\bx)$ of all vector fields associated to the finite states in $S$. If $m_\Kd>1$ then the Markov chain is equivalent  to a product of $m_\Kd$ Markov chains and the vector-field  (\ref{average}) describes the deterministic dynamics averaged over all $m_\Kd$ components of $S$. We illustrate the theory  using equation (\ref{average}) and derive as applications effective reaction rates related to different macro-molecular machinery. One prominent example is the well known Hill's kinetics. In a forthcoming paper we apply this theory to derive rigourously the nonlinear macroscopic model used in \cite{thattai} to study - on a more heuristic basis - the bistability in the \emph{Lac-Operon}.\\

The organisation of the paper is as follows. We first define systems with both infinite and finite degrees of freedom. Then  we introduce formally the continuum approximation for our setting. Finally the adiabatic approximation is constructed. The second part of this series contains the  examples, noteworthy a new approach to enzyme kinetics (\cite{sbano2}). 
In the appendix we describe the geometrical structure of the Markov chain which is very important to develop the adiabatic theory.

\section{Systems with infinite and finite degrees of freedom}
\label{SIFDF}

As motivated in the introduction typical macro-molecular systems give rise to mixed microscopic dynamics, consisting of birth-death processes where particle or molecule numbers can be arbitrarily large, and a second part where some entities have a fixed number of molecules in the system, but each equipped with finitely many different functional states giving rise to a finite state Markov chain.\\
 The following definition will fix this structure for further investigation, followed by an illustrative and biologically important example.
\bdefi
Let us define two sets of scales
\begin{enumerate}
\item size scales $\vde=(\de_1,...,\de_N)$, $\de_i>0$,
\item time scale $\tau>0$.
\end{enumerate}
Let $\LL_\vde$ be the following lattice
\begin{equation}
\label{lattice}
\LL_\vde\doteq\{\bn\,\vde=(n_1\,\de_1,...,n_N\,\de_N):~~\bn=(n_1,...,n_N)\in\N^N\}
\end{equation}
\edefi
\brem
On $\LL_\vde$ we shall define functions, now for fixed $\vde$ the value of any function $u$ 
is uniquely determined by the integer vector $\bn$  therefore whenever $\vde$ is fixed we shall omit the $\vde$ dependence and write $u(\bn)$.
\erem
\bdefi
Let the  tuple $(\zeta, R, P)$ determine a stochastic process by specifying the state $\zeta$, a set of reactions $R$, and a vector of probabilities $P$, such that

\begin{itemize}
\item[(i)] the state $\zeta$ of the system is fully specified by $n_1,...,n_N$ \emph{infinite degrees of freedom} (\emph{i.d.f.}) and a second variable, the \emph{finite degrees of freedom} $s$ (\emph{f.d.f.}). The state $\zeta$ is therefore the composition

\[\zeta=(n_1\de_1,....,n_N\de_N,\sig)=(\bn\vde,\sig) \in \LL_\vde\times \Sig,\]

where $\LL_\vde=\vde\N^N$, $\bn$ is an $n$-tuple of natural numbers and $\sig$ runs in a finite set $\Sig$, with $|\Sig|=g$ being the number of discrete states. 

\item[(ii)] the time evolution of the stochastic process is defined via the set of reactions $R$ having three different types:

\begin{enumerate}
\item[(a)] Processes involving only \idf's represented by reactions (possibly reversible) of the form

\[(\bn,\sig)\ra(\bn',\sig).\] 

The operator describing these reactions in the master equation (\ref{equation0})  is denoted by $\Ld^*_R$ and has the form
$\Ld^*_R=\ell_0\otimes\delta_{\sig\sig'j}$ where $\ell_0$ is the same operator for each discrete state $\sig=1,...,g$. Here $\delta_{\sig\sig'} =1$ for $\sig=\sig'$, and zero otherwise.

\item[(b)] Processes involving only \fdf's represented by 
reactions (possibly reversible) of the form

\[(\bn,\sig)\ra(\bn,\sig').\] 

The operator describing these reactions in the master equation (\ref{equation0})  is  the transpose $\Kd^T$ of  the Markov chain generator of  the process governing the transitions among the discrete states $\sig=1,...,g$.  The Markov chain is finite dimensional with a space of stationary states $M_\Kd$ of dimension strictly less than g.
 
\item[(c)] Processes involving both \idf~ and \fdf~  represented by 
reactions (possibly reversible) of the form

\[(\bn,\sig)\ra(\bn',\sig).\] 

The operator describing these reactions in the master equation (\ref{equation0}) is denoted by $\Ld^*_E$. This operator is non-trivial only in the discrete states $\sig$ which affect processes involving \idf.

\end{enumerate}
\end{itemize}

\item[(iii)] each realisation of the process is valued in $\LL_\vde^N\times \Sig$. The state $\zeta$ at time $t$ is given by the vector of probabilities

\begin{equation}
P(t,\bn)=(P_1(t,\bn),...,P_g(t,\bn)), \mbox{ with } \sum_{\bn\in\N^N}\sum_{\sig=1}^gP_\sig(t,\bn)=1. 
\label{e:problem}
\end{equation}

The time evolution of $P$ is given by the master equation (ME)

 \begin{equation}
 \frac{\pa P(t,\bn)}{\pa t}=(\Ld^*_R+\Ld^*_E)\circ P(t,\bn)+ \Kd^T(\bn)\,P(t,\bn), 
 \label{equation0}
 \end{equation}

$P$, $\Ld^*_R$, $\Ld^*_E$ and $\Kd^T$ are sufficiently regular such that (\ref{equation0}) has a unique solution for all times $t >0$. Then the tuple $(\zeta, R, P)$ is called a (microscopic) system with \emph{infinite and finite degrees of freedom}, or short an $\bf IFSS$ (Infinite-Finite State System).
\edefi

\subsubsection*{Illustrative example of a typical $\bf IFSS$: Single enzyme kinetics} Consider a system with 2 \idf~  and 1 \fdf. The system has state space $(a,m,O_i)\in(\de\N)^2\times S$, and will be described by the vector propbability 

\[   P(t,m,a)=(P_0(t,m,a),P_1(t,m,a)).  \] 

Here $a$ and $m$ are the numbers of two small species of molecules (\idf) called $A$ and $M$, and  $S=\{O_0,O_1\}$ are the discrete states of a molecular machine (\fdf), for example an enzyme.  Reactions of type $(a)$ are those independent from the discrete states of the macro-molecular machinery. In our example we assume that degradation of $M$ is of this type:

\[\begin{array}{ll}
M \rightarrow^{\nu/\tau}\,\emptyset  \\
\end{array}\]

This reaction takes place at speed $\nu > 0$. This gives the following contributions to the ME

\[+\frac{\nu}{\tau}\,(m+1)\,P_\sig(t,m+1,a)-\frac{\nu}{\tau}\,m\,P_\sig(t,m,a)~~\mbox{ for $\sig=O_0,O_1$}.\]

Such terms can be rewritten as

\[\frac{\nu}{\tau}\,(\Eop^+-\id)(m\,P_\sig(t,m,a))\]

where $\Eop^\pm$ and $\id$ are difference operator defined as 
\[\Eop^+f(m)=f(m+1),~~\id f(m)=f(m)\mbox{ for every $f:\de\N\ra\R$.}\]

\noi Using this difference operator notation $\Ld^*$ is then given by

\[
\Ld^*_R\, = \,\frac{1}{\tau} \left( \begin {array}{cc}
 \nu\,(\Eop^+-\id)(m\,\cdot \,) &0
 \\\noalign{\medskip}0 &  \nu\,(\Eop^+-\id)(m \,\cdot \,)
 \end {array} \right),
\]
 Reactions of type $(b)$ are given by 

\[\begin{array}{ll}
A+O_0 \rightarrow^{k^+/\tau} O_1, \\
O_1 \rightarrow^{k^-/\tau} O_0+A,
\end{array}\]

they describe the transitions of the discrete states of the macro-molecule, possibly depending on binding of smaller molecules, in this case of molecules of type $A$. 
Moreover the transition rates depend on the relative Markov chain 'switching' time scale $\tau$ and the system size $\delta$, which is defined as inverse of the largest average number of $A$ molecules in the system. The generator of the Markov chain is then given by

 \[
\Kd\, =  \frac{1}{\tau}\left( \begin {array}{cc} 
-a\,k^+ & a\,k^+
\\\noalign{\medskip}k^-& -k^-
\end {array} \right) .
\]

This also implies the time evolution of the system must be described through the vector probability

\[P(t,m,a)=(P_0(t,m,a),P_1(t,m,a-1))\simeq (P_0(t,m,a),P_1(t,m,a))\]
for $a$ much larger than $\de$.

Finally the only reaction of type $(c)$ is given by

\[\begin{array}{ll}
\emptyset\rightarrow^{v/\tau}\,M  \mbox{ for } \sig= O_1,
\end{array}\]

with  $O_1$ interpreted as the active state, the only one at which the enzyme in addition catalyses molecules of type $M$. This reaction gives the following contribution to the ME
\[\begin{array}{ll}
0\mbox{ for $s= O_0$}\\[3mm]
\displaystyle \frac{v}{\tau}\,P_1(t,m-1,a)-\frac{v}{\tau}\,P_1(t,m,a) \mbox{ for $\sig=O_1$.}
\end{array}\]
The second contribution can be rewritten as
\[\frac{\nu}{\tau}(\Eop^{-}-\id)P_\sig(t,m,a)\mbox{ for $\sig=O_1$ and $m\geq 1$.}\]
where $\Eop^{-}$ is the difference operator defined by 
\[\Eop^{-}(f(m))=f(m-1)\mbox{ for every $f:\de\N\ra\R$.}\]
The operator $\Ld_E^*$ is defined as

\[
\Ld_E^*\, = \frac{1}{\tau}\, \left( \begin {array}{cc}
 0 &0
 \\\noalign{\medskip}0 &  v\,(\Eop^- -\id)(\cdot \,)
 \end {array} \right) .
\]

The ME  can now be written as

\[\frac{\pa P}{\pa t}=\Ld^*\,P+\Kd^T\,P, \]

where $\Ld^*=\Ld_R^*+\Ld_E^*$. This is the matrix form of

\[\left\{\begin{array}{llll}
\displaystyle \frac{dP_0(t,m,a)}{dt}=\frac{\nu\,(m+1)}{\tau}\,P_0(t,m+1,a)-\frac{\nu\,m}{\tau}\,P_0(t,m,a)+\\[3mm]
\displaystyle-\frac{a\,k^+}{\tau}\,P_0(t,m,a)+\frac{k^-}{\tau}\,P_1(t,m,a)\\[4mm]
\displaystyle \frac{P_1(t,m,a)}{dt}=\frac{\nu\,(m+1)}{\tau}\,P_1(t,m+1,a)-\frac{\nu\,m}{\tau}\,P_1(t,m,a)+\\[3mm]
\displaystyle+\frac{v}{\tau}\,P_1(t,m-1,a)-\frac{v}{\tau}\,P_1(t,m,a)+\frac{a\,k^+}{\tau}\,P_0(t,m,a)-k^-\,P_1(t,m,a)
\end{array}\right.\]

\noi This system will be fully analysed in the second part, see \cite{sbano2}. 
The boundary conditions are the natural one at $m=0$ (see \cite{Vankampen}) and are give by
\[\left\{\begin{array}{ll}
\displaystyle \frac{dP_0(t,0,a)}{dt}=\frac{\nu}{\tau}\,P_0(t,1,a)-\frac{a\,k^+}{\tau}\,P_0(t,0,a)+\frac{k^-}{\tau}\,P_1(t,0,a)\\[4mm]
\displaystyle \frac{P_1(t,0,a)}{dt}=\frac{\nu}{\tau}\,P_1(t,1,a)+\frac{a\,k^+}{\tau}\,P_0(t,0,a)-k^-\,P_1(t,0,a)
\end{array}\right.\]


\section{Construction of the continuum approximation}
\label{continuum}
The ME results from the specification of the reactions at a given scales $\vde,\tau$. 
The ME describes the evolution of a probability measure $P_\sig(t;\bn)$ according to
\begin{equation}
\frac{\pa P}{\pa t}=\A^*[\vde,\tau]P,
\label{eq-adjA}
\end{equation}
where $\A^*[\vde,\tau]$ is the infinitesimal generator defined on the scales $\vde,\tau$ by
\begin{equation}
\A^*[\vde,\tau]\doteq\Ld^*[\vde,\tau]+\Kd^T[\vde,\tau].
\label{op-A}
\end{equation}
The operator $\A^*[\vde,\tau]$ is defined on the space 
\begin{equation}
\X^*_{\vde,\tau}\doteq\left\{P_\sig(t;\bn):\sum_{\bn\in\LL_\de}\sum_{\sigma\in\Sigma}
P_\sig(t;\bn)=1\mbox{  for al $t$}\right\}.
\label{dualXn}
\end{equation}
Let us now consider a sequence of scales $\vde_n,\tau_n$ such that $\vde_n\ra 0$ 
and $\tau_n\ra 0$ as $n\ra \infty$. For each index $n$ we have an operator $\A_n=\A[\vde_n,\tau_n]$ defined on $\X^*_n=\X^*_{\vde_n,\tau_n}$ where the configuration space can now be denoted by $\LL_n=\LL_{\vde_n}$. We ask ourselves what would the fate of (\ref{op-A}) be as $n\ra\infty$.\\
We can think of $\vde_n\ra 0$ and $\tau_n\ra 0$  as limit at which space and time step 
become continuous and the numbers of particles are sufficiently to be accounted as densities and this motivates the name \emph{continuum limit}.\\
The formulation of the continuum limit can be obtained  by using the approximation 
scheme introduced by Trotter in \cite{Trotter}, (see also \cite{Pazy}, \cite{Kurtz}). 
To introduce Trotter approach we first need to observe that to each $\A^*_n$ defined on $\X^*_n$ we can associate a vector space $\X_n$ and an adjoint operator $\A_n$.
The vector space is defined by
\begin{equation}
\label{Xspace}
\X_{n}=\X_{\vde_n,\tau_n}\doteq\left\{u_\sig(t,\bn):\LL_{n}\times\Sigma\ra\R^g:
\|u\|_\infty=\sup_{\bn\in\LL_\de,\sigma\in\Sigma}|u_\sig(t,\bn)|<\infty, \mbox{  for al $t$}\right\}.
\end{equation}
Each $\X_n$ is dual to $\X^*_n$ according to the pairing:
\begin{equation}
\label{nduality}
\langle u,P\rangle_n\doteq\sum_{(\bn,\sigma)\in\LL_n\times\Sigma}u_\sig(\bn)\,P_\sig(\bn).
\end{equation}
The adjoint $\A_n$ is defined by:
\begin{equation}
\langle\A_nu,P\rangle_n=\langle u,\A^*_nP\rangle_n.
\end{equation}
Let us consider
\[u(t,\bn,\sig)=\sum_{\bn',\sig'}P(t,\bn,\bn',\sig,\sig')u(\bn',\sig')\]
then to equation (\ref{eq-adjA}) we now associate
 \begin{equation}
\frac{\pa u}{\pa t}=\A_n\,u,
\label{eq-A}
\end{equation}
defined on each $\X_n$. Here
\begin{equation}
\A_n\doteq\lim_{t\ra 0}\frac{1}{t}(P^t-\id)
\end{equation}
see \cite{Wentzell} for all the details.\\
For any index $n$ equation (\ref{eq-A}) is the standard 
Kolmogorov and $\A_n$ is the infinitesimal generator of Markov process 
on $\LL_n\times\Sigma$. 
The definition of the continuum limit is based on the choice a target space 
where the limit is attained. We shall consider as target the space of 
continuous function $\X=C^0(\R_+^N,\R^g)$.  The topological dual of $\X$ is formed 
by signed measures on $\R^N\times\Sigma$:
\begin{equation}
\X^*=\left\{\rho(\bx):\langle\rho,u\rangle<\infty,~~u\in\X\right\},
\end{equation}
where the pairing is defined by 
\[\langle\rho,u\rangle\doteq \int_{\R_+^N}d\bx\,\sum_{\sigma\in\Sigma}\rho_\sigma(\bx)
u_\sigma(\bx).\]
According to \cite{Trotter} we define a sequence of projections
\bdefi
Let $\Pro_n:\X\mapsto\X_n$ be the operator that maps $u\in\X$ to
$\Pro_n(u)\in\X_n$ defined as
\[\Pro_n(u)(\bk)=u(\bk\,\vde_n)=u(k_1\de^1_n,...,k_N\de_n^N).\]
\edefi
The following holds true
\bprop
The projections $\Pro_n$ satisfy the following properties
\begin{itemize}
\item[(i)] $\|\Pro_n\|_n<1$,
\item[(ii)] $\lim_{n\ra\infty}\|\Pro_n(u)\|_n=\|u\|_\infty$ for every $u\in\X$.
\end{itemize}
\eprop
\bpf
Part (i) follows from:
\[\|\Pro_n\|_n=\sup_{u\in\X}\frac{\|\Pro_n(u)\|_n}{\|u\|_\infty}\]
and from 
\[\|\Pro_n(u)\|_n=\sup_{\bk\in\LL_n}\|u(\bk\,\vde_n)\|<\|u\|_\infty\]
for $\LL_n\subset\LL_{n+1}$.\\
Let us now show (ii). We know that $\|\Pro_n(u)\|_n<\|u\|_\infty$. Since 
$\LL_n\subset\LL_{n+1}$ the following holds
\[\|\Pro_n(u)\|_n\leq\|\Pro_{n+1}(u)\|_{n+1}.\]
therefore the sequence $\|\Pro_n(u)\|_n$ is increasing and as $n\ra\infty$
\[\|\Pro_n(u)\|_n\ra\sup_n\|\Pro_n(u)\|_n=\|u\|_\infty.\]
\epf
Following (\cite{Trotter}) the projectors $\Pro_n$ allow to define in what sense 
the spaces $\X_n$ approximate $\X$.

\bdefi
A sequence $u_n\in\X_n$ converges to $u\in\X$ if 
\[\|\Pro_n(u)-u_n\|_n\ra 0\mbox{ as $n\ra\infty$}.\]
We denote this by $u_n\app u$. 
\edefi
\brem
Condition (ii) on $\Pro_n$ guarantees that the limit $\app$ is unique. In fact if 
$u_n\app u$ and $u_n\app u'$ then by (ii) we can estimate
\[\|u-u'\|_\infty=\lim_{n\ra\infty}\|\Pro_n(u-u')\|_n\leq\lim_{n\ra\infty}\|\Pro_n(u)-u_n\| +
\lim_{n\ra\infty}\|\Pro_n(u')-u_n\|_n\]
which of course goes to $0$ as $n\ra\infty$.
\erem
We now give the definition for the limit, in fact the \emph{continuum limit}, 
of a sequence of operators $\A_n$. This definition is inspired by the one presented in \cite{Trotter}. In fact in the present case we have to consider that the operators are functions of the scales $\vde_n$ and $\tau_n$. Therefore we set
\bdefi
Let $\A_n:\X_n\mapsto\X_n$ be a sequence of linear operators. We say that 
$\Aop:\X\mapsto\X$ is the continuum limit of $\A_n$ (denoted by $\A_n\app\Aop$) if there exists a sequence of scales $\vde_n,\tau_n$ such that
\begin{enumerate}
\item $\vde_n\ra o$, $\tau_n\ra 0$,
\item the domain of $\A$ is 
\[D(\Aop)=\{u\in\X:~\Pro_n(u)\in D(\A_n),~~ \A_n(\Pro_n(u))\mbox{ converges}\},\]
\item and $\|\Pro_n(\Aop(u))-\A_n(\Pro_n(u))\|_n\ra 0$ as $n\ra\infty$.
\end{enumerate}
\edefi

\brem
The dependence on the choice of the scales $\vde_n$ and $\tau_n$ makes the continuum limit non unique.  This is very important because with the choice of the scaling 
we will be able to analyse different type of processes.  
\erem

\begin{figure}[htbp] 
   \centering
   \includegraphics[scale=0.6]{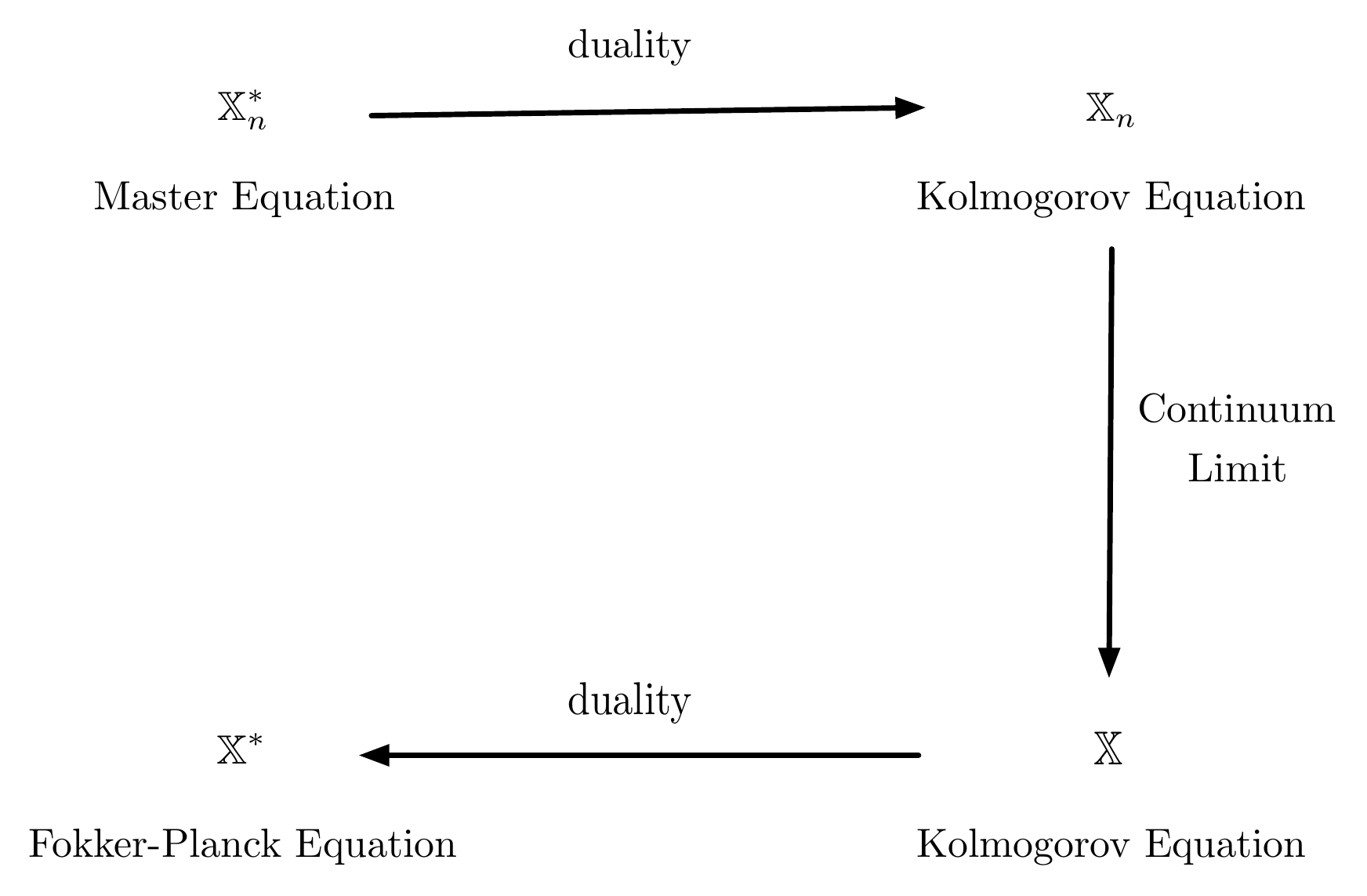} 
   \caption{\small A schematic view of the Continuum Limit }
   \label{fig:continuum}
\end{figure}

In what follows we shall construct examples $\A^*_n$ that correspond to Master Equations on $\X_n^*$. Via the duality we derive the infinitesimal generator 
$\A_n$ i.e. the Kolmogorv equation on $\X_n$. At that point we study the continuum limit 
of $\A_n$ and finally via the duality between $\X^*$ and $\X$ we derive the Fokker-Planck equation on $\X^*$. 
This is illustrated in Figure \ref{fig:continuum}.

\subsection{Examples}
In this section we illustrate the theory by studying the continuum limit of 
some crucial examples that will be used in the applications to reaction networks.\\
\subsubsection{Difference operator}
\label{diff-op}
Let $\LL_n=\de_n\,\Z$ and $X^*_n$ be the space of probability distribution on $\LL_n$ and $\X=C^0(\R,\R)$. Consider the following operator
\[\Dop^*_n(P)(k)=\frac{1}{\tau_n}(\Eop^+-\id)(a_nP)(k)=\frac{1}{\tau_n}[a_n(k+1)P(k+1)-a_n(k)P(k)]\]
The adjoint $\Dop_n$ is defined by
\[\lb \Dop^*_nP,u\rb_n=\lb P,\Dop_n u\rb.\]
A simple calculation shows that
\[\Dop_n=\frac{a_n(k)}{\tau_n}(\Eop^--\id).\]
We now compute the continuum limit. Let $a\in\X$ such that $\Pro_n(a)(k\de_n)
=a(k\de_n)=a_n(k)$. Let $\de_n,\tau_n$ such that 
\[\de_n\ra 0,\tau_n\ra 0\mbox{ with } \frac{\de_n}{\tau_n}\ra c>0\mbox{ as $n\ra\infty$}.\]
Then 
\[\Dop_n\app-c a(x)\Dop,~\mbox{ where } \Dop=\frac{\pa}{\pa x}.\]
In fact consider take $u\in C^2(\R,\R)\subset \X$
\[\|\Dop_n(\Pro_n(u))-\Pro_n(\Dop(u))\|_n=\left\|\frac{a_n(k)}{\tau_n}(\Eop^--\id)u(k\,\de_n)+
c\,a(k\,\de_n)\frac{\pa u}{\pa x}(k\,\de_n)\right\|_n.\]
This can be rewritten as
\[\begin{array}{lll}
\displaystyle\left\|\frac{a_n(k)\,\de_n}{\tau_n}\left(\frac{u((k-1)\,\de_n)-u(k\,\de_n)}{\de_n}\right)+
c\,a(k\,\de_n)\frac{\pa u}{\pa x}(k\,\de_n)\right\|_n=\\[4mm]
\displaystyle =\left\|a_n(k)\,\left(\frac{\de_n}{\tau_n}-c\right)\left(\frac{u((k-1)\,\de_n)-u(k\,\de_n)}{\de_n}\right)+
c\,\left(a(k\,\de_n)\frac{\pa u}{\pa x}(k\,\de_n)+a_n(k)\frac{u((k-1)\,\de_n)-u(k\,\de_n)}{\de_n}\right)\right\|_n=\\[4mm]
\displaystyle=\left\|a(k\,\de_n)\,\left(\frac{\de_n}{\tau_n}-c\right)\left(\frac{u((k-1)\,\de_n)-u(k\,\de_n)}{\de_n}\right)+
c\,a(k\,\de_n)\,\left(\frac{\pa u}{\pa x}(k\,\de_n)+\frac{u((k-1)\,\de_n)-u(k\,\de_n)}{\de_n}\right)\right\|_n
\end{array}\]
The last term is bounded by
\[\begin{array}{ll}
\displaystyle\left\|a(k\,\de_n)\,\left(\frac{\de_n}{\tau_n}-c\right)\left(\frac{u((k-1)\,\de_n)-u(k\,\de_n)}{\de_n}\right)+
c\,a(k\,\de_n)\,\left(\frac{\pa u}{\pa x}(k\,\de_n)+\frac{u((k-1)\,\de_n)-u(k\,\de_n)}{\de_n}\right)\right\|_n\leq\\[4mm]
\leq \displaystyle\left|\frac{\de_n}{\tau_n}-c\right|\sup_k|a(k\,\de_n)|
\sup_k\left|  \frac{u((k-1)\,\de_n)-u(k\,\de_n)}{\de_n}\right| +c\sup_k|a(k\,\de_n)|\sup_k\left| \frac{\pa u}{\pa x}(k\,\de_n)+ \frac{u((k-1)\,\de_n)-u(k\,\de_n)}{\de_n}\right|.
\end{array}\]
Now as $n\ra\infty$ 
\[\left|\frac{\de_n}{\tau_n}-c\right|\ra 0,\]
$\sup_k|a(k\,\de_n)|$ is bounded, the term
\[\sup_k\left|  \frac{u((k-1)\,\de_n)-u(k\,\de_n)}{\de_n}\right|=
\sup_k\frac{1}{\de_n}\left|\int_{(k-1)\de_n}^{k\,\de_n}dx\frac{u(x)}{\pa x}\right|\leq
\left\|\frac{u(x)}{\pa x}\right\|_\infty.\]
For the term
\[\sup_k\left| \frac{\pa u}{\pa x}(k\,\de_n)+ \frac{u((k-1)\,\de_n)-u(k\,\de_n)}{\de_n}\right|\]
we use
\[u(y)=u(x)+u'(x)(y-x)+\int_x^yds(s-x)\frac{\pa^2 u(s)}{\pa s^2}\]
to obtain
\[\begin{array}{lll}
\displaystyle\sup_k\left| \frac{\pa u}{\pa x}(k\,\de_n)+ \frac{u((k-1)\,\de_n)-u(k\,\de_n)}{\de_n}\right|=\\[4mm]
=\displaystyle\sup_k\left|\frac{1}{\de_n}\int_{(k-1)\de_n}^{k\,\de_n}ds\,(s-k\,\de_n)\frac{\pa^2 u(s)}{\pa s^2}\right|\leq\\[4mm]
\displaystyle\leq\|u''\|_\infty\frac{1}{\de_n}\sup_k\left|\int_{(k-1)\de_n}^{k\,\de_n}ds\,(s-k\,\de_n)\right|\leq\|u''\|_\infty\frac{\de_n}{2}\ra 0\mbox{ as $n\ra\infty$.}
\end{array}\] 
We therefore conclude $\A_n\app\Aop$. A  simple integration by parts shows that 
\[\Dop^*(\rho)(x)=\frac{\pa}{\pa x}(c\,a(x)\,\rho(x))\mbox{ for all $\rho\in\X^*$}.\]

\subsubsection{Multiplication operator}
Let $\Kd_n=\Kd_{\de_n,\tau_n}(\bn)$ be defined on $\X_n$. We want to find $
\Kop$ is a matrix operator acting by multiplication on $\X$ such that
\[\|\Kd_n(\Pro_n(u))-\Pro_n(\Kop(u))\|_n\ra 0~~\forall u\in\X.\]
This condition holds if we take $\Kop(\bx)=\frac{1}{\veps}\,\Kd(\bx)$ such that
\[\left\|\Kd_{\de_n,\tau_n}(\bn)u(\bk\,\vde_n)-\frac{1}{\veps}\Kop(u(\bk\,\vde_n)))\right\|_n\ra 0~~\forall u\in\X,\]
where $\veps$ is chosen so that the limits holds.\\
Let us consider the following example 
\[\Kd_n=\frac{1}{\tau_n}\left(\begin{array}{cc}
-m\,k^+(\de_n,\tau_n) &~~ m\,k^+(\de_n,\tau_n)\\
k^-(\de_n,\tau_n) &~~ -k^-(\de_n,\tau_n)
\end{array}\right).\]
We seek the continuum limit in the form
\[\Kd=\frac{1}{\eps}\left(\begin{array}{cc}
-x\,k^+ &~~ x\,k^+\\
k^- & -k^-
\end{array}\right)\]
with $k^\pm$ positive constants. 
Now in order to construct the expression
\[\|\Kd_n\Pro_n(u)-\Pro_n(\Kd(u))\|_n\]
we compute
\[\Kd_n(\Pro_n(u))(m)=\frac{1}{\tau_n}\left(\begin{array}{c}
-m\,k^+(\de_n,\tau_n)(u_1(m\,\de_n)-u_1(m\,\de_n))\\[2mm]
k^-(\de_n,\tau_n)(u_1(m\,\de_n)-u_1(m\,\de_n))
\end{array}\right)\]
and
\[\Pro_n(\Kd(u))(m)=\frac{1}{\eps}\left(\begin{array}{c}
-m\,\de_n\,k^+(\de_n,\tau_n)(u_1(m\,\de_n)-u_1(m\,\de_n))\\[2mm]
k^-(\de_n,\tau_n)(u_1(m\,\de_n)-u_1(m\,\de_n))
\end{array}\right).\]
Therefore we can have
\[\lim_{n\ra\infty}\|\Kd_n\Pro_n(u)-\Pro_n(\Kd(u))\|_n= 0\]
for all $u\in\X$ if the following relations hold
\[\frac{\de_nk^+}{\eps}=\frac{k^+(\de_n,\tau_n)}{\tau_n},~~
\frac{k^-}{\eps}=\frac{k^-(\de_n,\tau_n)}{\tau_n}.\]
These relations have to satisfy the compatibility condition 
\[\frac{\de_n\,k^+}{k^-}=\frac{k^+(\de_n,\tau_n)}{k^-(\de_n,\tau_n)}.\]
Here we see that if
\[k^+(\de_n,\tau_n)=\de_n\,k^+,~~k^-(\de_n,\tau_n)=k^-\]
then the limit exits for $\eps=\tau_n$ which is infinitesimal for large $n$.
\brem
The choice of $\eps$ as a function of of the scales $\vde_n,\tau_n$ 
is essentially a form a form of renormalisation of the  
time scales  associated to respectively the infinitesimal generator $\Lop$ and 
$\Kop$.
\erem

\subsubsection{Diffusion operator}
WE consider the one dimensional case $\LL_n=\de_n\Z$. Let us consider the operator
\begin{equation}
\label{diffusionOP}
\A_n=\frac{1}{\tau_n}[a_n(k)\Dop^-_n+b_n(k)\Dop^+_n]
\end{equation}
We now take $a,b\in\X$ such that $\Pro_n(a)=a_n$ and $\Pro_n(b)=b_n$.
In order to construct a continuum limit we compute
\[\begin{array}{ll}
\displaystyle\A_n(\Pro_n(u))=\frac{1}{\tau_n}a_n\Dop^-_n(u)+\frac{1}{\tau_n}b_n\Dop^+_n(u)=\\[4mm]
\displaystyle=\frac{a_n(k)}{\tau_n}[u((k-1)\de_n)-u(k\de_n)]+
\frac{b_n(k)}{\tau_n}[u((k+1)\de_n)-u(k\de_n)].
\end{array}\]
We now re-write this operator using $u\in C^3(\R,\R)$ using:
\[u(y)=u(x)+u'(x)(y-x)+\frac{(y-x)^2}{2}u''(x)+\frac{1}{6}\int_x^yds(s-x)^2\frac{\pa^3 u(s)}{\pa s^3}.\]
We have
\[\begin{array}{lll}
\displaystyle\A_n(\Pro_n(u))=\frac{a_n(k)}{\tau_n}\left[-\de_n\frac{\pa u(k\de_n)}{\pa x}+
\frac{\de_n^2}{2}\frac{\pa^2 u(k\de_n)}{\pa x^2}\right]+
\frac{b_n(k)}{\tau_n}\left[\de_n\frac{\pa u(k\de_n)}{\pa x}+
\frac{\de_n^2}{2}\frac{\pa^2 u(k\de_n)}{\pa x^2}\right]+\\[4mm]
\displaystyle+\frac{a_n(k)}{6\tau_n}\int_{k\de_n}^{(k-1)\de_n}ds(s-k\de_n)^2\frac{\pa^3 u(s)}{\pa s^3}
+\frac{b_n(k)}{6\tau_n}\int^{(k+1)\de_n}_{k\de_n}ds(s-k\de_n)^2\frac{\pa^3 u(s)}{\pa s^3},
\end{array}\]
this form can be rewritten as
\[\begin{array}{lll}
\displaystyle\A_n(\Pro_n(u))=\frac{\de_n(b_n(k)-a_n(k))}{\tau_n}\frac{\pa u(k\de_n)}{\pa x}+\frac{\de_n^2(b_n(k)+a_n(k))}{2\tau_n}\frac{\pa^2 u(k\de_n)}{\pa x^2}+\\[4mm]
\displaystyle
+\frac{a_n(k)}{6\tau_n}\int_{k\de_n}^{(k-1)\de_n}ds(s-k\de_n)^2\frac{\pa^3 u(s)}{\pa s^3}+
\frac{b_n(k)}{6\tau_n}\int^{(k+1)\de_n}_{k\de_n}ds(s-k\de_n)^2\frac{\pa^3 u(s)}{\pa s^3},
\end{array}\]

\noi Now we need to guess a suitable operator $\A$. Here we take
\[\Aop=\alpha(x)\frac{\pa}{\pa x}+\beta(x)\frac{\pa^2}{\pa x^2}\]
with $\alpha,\beta \in\X$. Therefore
\[\Aop(\Pro_n(u))(k)=\alpha(k\de_n)\frac{\pa u(k\de_n)}{\pa x}+\beta(k\de_n)\frac{\pa^2u(k\de_n)}{\pa x^2}.\]
If we set $\de_n^2/\tau_n\ra 1$ and 
\[\|\alpha(k\de_n)-\frac{\de_n}{\tau_n}(b_n(k)-a_n(k))\|_n=o(1/n),~~\|\beta(k\de_n)-\frac{\de_n^2}{2\tau_n}(b_n(k)+a_n(k))\|_n=o(1/n)\]
then
\[\|\A_n(\Pro_n(u))-\Pro_n(\A(u))\|_n\leq o(1/n)(\|u'\|_\infty+\|u''\|_\infty)+\|R(\de_n,\tau_n)\|_n,\]
where
\[R(\de_n,\tau_n)=\frac{a_n(k)}{6\tau_n}\int_{k\de_n}^{(k-1)\de_n}ds(s-k\de_n)^2\frac{\pa^3 u(s)}{\pa s^3}+
\frac{b_n(k)}{6\tau_n}\int_{(k+1)\de_n}^{k\de_n}ds(s-k\de_n)^2\frac{\pa^3 u(s)}{\pa s^3}.\]
Now we can estimate
\[\|R(\de_n,\tau_n)\|_n\leq\|a_n+b_n\|_n\left\|\frac{\pa^3 u}{\pa s^3}\right\|_\infty\frac{\de_n^3}{6\tau_n}\sup_k\left|\frac{1}{3}(k-1-k)^3+\frac{1}{3}(k+1-k)^3\right|
\leq\|a_n+b_n\|_n\left\|\frac{\pa^3 u}{\pa s^3}\right\|_\infty\frac{\de_n^3}{9\tau_n}\]
Note that $\|a_n+b_n\|_n\ra \|\alpha+\beta\|_\infty$ and thus have $\|R(\de_n,\tau_n)\|_n\ra 0$ as $n\ra\infty$ .\\
Also in this case using the paring between $\X$ and $\X^*$ we can show by an integration by parts that 
\[\Aop^*(\rho(x))=-\frac{\pa}{\pa x}(\alpha(x)\rho(x))+\frac{\pa^2}{\pa x^2}(\beta(x)\rho(x))\]
for all $\rho\in\X^*$.

 \subsection{Formulation of the multi-scale analysis}
 \label{multiscale}

We call a {\it multiscale analysis} the combination of continuum approximation and an adiabatic approximation 
for the ME of an IFSS. We construct the continuum approximation of the birth-death processes describing the small molecules first  and will get a Fokker-Planck equation (FPE). Next we study the adiabatic approximation of the Markov chain, ending the multi-scale analysis.

 \subsubsection{Continuum limit of the ME}

 \bdefi[Continuum approximation]  Consider the infinitesimal generator associated to ME of an IFSS given by
 \begin{equation}
 \label{equation1}
 \frac{\pa u(\bn,t)}{\pa t}=\Ld[\vde,\tau](\bn) u(\bn,t)+\Kd[\vde,\tau](\bn)\,u(\bn,t),
 \end{equation}
\noindent defined on $\X_\vde$. We say that (\ref{equation1}) admits as continuum limit
 \begin{equation}
 \label{continuum-equation1}
 \frac{\pa u(\bx,t)}{\pa t}=\Lop(\bx) u(\bx,t)+\frac{1}{\eps}\,\Kop(\bx)\,u(\bx,t),
 \end{equation}
\noindent defined on $\X$ if there exist a choice of the scaling $\vde_n\ra 0,\tau_n\ra 0$ as $n\ra\infty$ such that 
\begin{itemize}
\item[(i)] $\Ld_n=\Ld[\vde_n,\tau_n]\app\Lop$, 
\item[(ii)]  $\Kd_n=\Kd[\vde_n,\tau_n]\app\frac{1}{\eps}\,\Kop$ for some choice of $\eps$.
\end{itemize}
\edefi

\begin{equation}
\frac{\pa \rho(\bx,\tau)}{\pa \tau}=\Lop^*(\bx)\rho(\bx,\tau)+\frac{1}{\eps}\,K^T(\bx)\,\rho(\bx,\tau).
\label{main-ME-K}
\end{equation}

\brem
Clearly equation (\ref{main-ME-K}) depends on the scales $\vde_n,\tau_n$ through 
the parameter $\eps$.
\erem

\subsubsection*{Illustrative example: continuum limit}
In the illustrative example from enzyme kinetics the matrix $\Ld^*$ is
\[
\Ld=\Ld_E+\Ld_R\, = \,\frac{1}{\tau} \left( \begin {array}{cc}
 \nu\,m\,(\Eop^--\id)(\cdot \,) &0
 \\\noalign{\medskip}0 &  \nu\,m\,(\Eop^--\id)(\cdot \,)+v\,(\Eop^+ -\id)(\cdot \,)
 \end {array} \right),
\]
 and
 \[
\Kd\, =  \frac{1}{\tau}\left( \begin {array}{cc} 
-a\,k^+& a\,k^+
\\\noalign{\medskip}k^- & -k^-
\end {array} \right) .
\]
Assuming that we can find $\eps=\eps(\de,\tau)$ such that 
\[(\de\nu(\de,\tau)/\tau)=\nu,~~(\de\,v(\de,\tau)/\tau)=v,~~(a\,k^+(\de,\tau)/\tau)=(a\,\de\,k^+)/\eps,~~(k^-(\de,\tau)/\tau)= k^-/\eps.\]
Taking $\eps(\de,\tau)=\tau$ with $\de/\tau\ra 1$, by recalling section \ref{diff-op} we obtain

\[
\Lop= \, \left( \begin {array}{cc}
 -\nu\,x\,\Dop(\cdot)&0
 \\\noalign{\medskip}0 &  -(\nu\,x-v)\Dop(\cdot)
 \end {array} \right),
\]

where $x$ is the concentration associated to $m$ and 

 \[
\Kd\, = \frac{1}{\eps} \left( \begin {array}{cc} 
-a\,k^+ & a\,k^+
\\\noalign{\medskip}k^-& -k^-
\end {array} \right) .
\]

The continuum limit provides the 
 construction of the infinitesimal operator $\Aop$
\[\Aop(\bx)=\Lop(\bx)+\frac{1}{\eps}\,\Kd(\bx)\]
densily defined in $\X$ and the associated Kolmogorov equation 
\begin{equation}
\frac{\pa u(\bx,t)}{\pa t}=\Aop(\bx)u(\bx,t).
\label{main-kolmogorov}
\end{equation}
 For this equation there is an adjoint formulation. This is the FPE that is 
 the dynamics read on the the space of probability measures $\X^*$. The FPE is
\begin{equation}
\frac{\pa \rho(\bx,\tau)}{\pa \tau}=\Aop^*(\bx)\rho(\bx,t),
\label{main-ME}
\end{equation}
where $\Aop^*$ is the adjoint of $\Aop$
\[\Aop^*=\Lop^*(\bx)+\frac{1}{\eps}\,\Kd^T(\bx).\]
 In most of the application the interests is concentrated on the the FPE.  
For this reason, in the next section we shall study the \emph{adiabatic limit} 
for (\ref{main-ME}). The dynamics on $\X$ is related to the dynamics on $\X^*$ by the following 
 the following result (see \cite{Kato,Wentzell}):
 \bth 
 Let  $\X$ be the Banach space of continuous functions from $\R^{N}$ to 
 $(\R^g,\langle.,.\rangle)$. For $\eps>0$ fixed, assume
 
 \begin{itemize}
 \item[(i)] $\Lop(\bx)$ has a dense domain in $\X$ and 
 
 \[\|(\Lop(\bx)+\alpha)^{-1}\|<\frac{1}{\alpha}~~\forall\alpha>0,\]
 
 where $\|.\||$ is the standard norm on the space of linear operators in $C^0(\R_+^{N},\R^g)$.
 
 \item[(ii)] $\Kd(\bx)$ is a $g\times g$ matrix with bounded entries, and also the infinitesimal generator of a 
 $g$-dimensional Markov chain.
 
 \end{itemize}
 
\noi Then equation 
 
 \begin{equation}
\frac{\pa u(\bx,t)}{\pa t}=\Lop(\bx)(u(\bx,t))+\frac{1}{\eps}\,\Kd(\bx)\,u(\bx,t)
\label{main-new-ME}
\end{equation}   

\noi admits a solution $u(\bx,t)$, and the operator $\Lop(\bx)+\frac{1}{\eps}\,\Kd(\bx)$ generates a Markov process on $\R^N\times S$ whose distribution $\rho(\bx,t)$ satisfies equation 
(\ref{main-ME}).
 
 \eth

As we have seen in the example $\eps$ is small, therefore FPE (\ref{main-ME}) for $\rho$ is singular at $\eps=0$. For this reason we shall consider a perturbation analysis by means of an asymptotic series in $\eps$ in the spirit of  \cite{pavliotis}. This will require a time-scale analysis of the FPE which is called \emph{adiabatic approximation}. We will need the following definitions:

\bdefi
Let $U$ be an open and bounded set in $\R^N_+$, $T>0$, and let $C^{r,s}$ be the Banach space of functions from $U\times [0,T]\subset\R_+^N\times\R$ to $\R^g$ 
which are $r$-times continuously differentiable w.r.t. $\bx\in U\subset\R^N$ and 
$s$-times continuously differentiable w.r.t. $t\in[0,T]\subset\R$. 
\edefi

We need to recall the definition of asymptotic series:

\bdefi[Asymptotic series]
Let $v_\eps(\bx,t)$ be the formal power series

\[v_\eps(\bx,t)=\sum_{k=0}^\infty v^{(k)}(\bx,t)\,\eps^k.\]

Then we say $v_\eps$ converges asymptotically to $v_0$ for small $\eps >0$ if its partial sum

\[v^m_\eps(\bx,t)=\sum_{k=0}^m v^{(k)}(\bx,t)\,\eps^k\]

is such that

\[\sup_{\bx\in U\subset \R^N,t\in [0,T]}\|v^m_\eps(\bx,t)-v_\eps(\bx,t)\|\leq C(U,T)\eps^{m+1}\]

for some $C(U,T)>0$.
\edefi

\section{Adiabatic theory}

In this section we construct the solution of the FPE generated by an IFSS by means of asymptotic expansions. Only the respective result for the  Kolmogorov equation will be stated. This choice is not generic from a purely mathematical point of view as a solution of the FPE will the require more regularity conditions. Nevertheless  the study of the FPE with smooth coefficients and smooth initial data is very relevant in many applications, motivating the subsequent presentation.

\subsection{Formulation of the adiabatic problem}

An IFSS has by necessity two time scales after taking the continuum limit. One is characterising the dynamics of the continuous degrees of freedom  and the other one characterising the evolution of the finite state Markov chain. These time scales are associated to the following semi-groups:

\begin{itemize}
\item[(i)] $\exp(t\,\Lop(\bx))$, the semigroup generated by $\Lop(\bx)$, 
\item[(ii)] $\exp((t/\eps)\,\Kd(\bx))$, which is the semigroup generated by $\Kd(\bx)$, the Markov chain generator.
\end{itemize}

\noi On a time scale of order $O(\eps)$ the Markov chain dynamics should prevail. Without loss of generality we can assume that the Markov chain has at least one invariant measure, and possibly a convex combination of stationary measures. Therefore on long time scales one expects that the Markov chain reaches an equilibrium very fast and the dynamics should essentially be given by the flow associated to $\Lop(\bx)$.

\subsubsection{Diffusive and deterministic operators}

The operators $\Lop(\bx)$ and $\Lop^*(\bx)$ are  differential operators which we investigate in the following, especially giving more  details of their structure. The operator  $\Lop^*(\bx)$ is a diagonal matrix  with operator entries. Each non-degenerate nonzero entry is a second order linear parabolic operator. 

\bdefi[Structure of $\Lop^*$] Let

\[\Lop^*(\bx)\doteq\Lop^1(\bx)+\Lop^2(\bx)=\delta_{ij}\otimes(
\Lop^1_i(\bx)+\Lop^2_i(\bx))\]

\noi Let each of these operators  $\Lop^1(\bx)$ and $\,\Lop^2(\bx)$ be given by

\[\Lop^1_i(\bx)(f)=\sum^N_\alpha\frac{\pa}{\pa x_\alpha}(L^j_\alpha(\bx)\,f(\bx)),  \]

and

\[\Lop^2_i(\bx)(f)=\frac{1}{2}\sum^N_{\alpha,\beta}\frac{\pa^2}{\pa x_\alpha\pa x_\beta}(C^i_{\alpha\beta}(\bx)\,f(\bx)),\]
\label{def:FP}

for $f$ being a sufficiently smooth function. Then $\Lop^*(\bx)$ is called a regular Fokker-Planck operator.
\edefi

If $\Lop^2\equiv 0$, the operator $\Lop^*$ becomes first order (i.e. non-regular) and describes the transport in the deterministic dynamics of \idf's. We can therefore identify two important regimes: With  $\Lop^2\neq 0$ we have diffusive \idf's, and with $\Lop^2\equiv 0$ the \idf's are deterministic. The second possibility will be treated in part II of this paper series.

\subsubsection{Main assumptions}
In order to simplify the further analysis we make the following assumption:

\begin{itemize}
\item[(A)] The Markov chain on $\Sig$ has a set of stationary measures $M_\Kd$ with $\dim(M_\Kd)<g$. Each measure $\mu(\bx)$ is $C^\infty$ on $\R^N$.
\end{itemize}

\noi The next two assumptions give the explicit conditions for constructing the solution, respectively for the FPE and the Kolmogorov equation.

\begin{itemize}
\item[(B)] For a given $\mu(\bx)\in C_\Kd$, the Cauchy problem

\begin{equation}
\pa_tf(\bx,t)=\langle\be_\mu,\Lop^*(\bx)(\mu(\bx) f(\bx,t))\rangle +F(\bx,t),~~~f(\bx,0)\in C^{r,s},
\label{eqB}
\end{equation}

with $F\in C^{r,s}$ admits a solution which is $C^{r,s}$ w.r.t. $\bx$ and $t\in [0,T_0]\subset [0,T]$ for any smooth initial data. 

\item[(C)]
For a given $\mu(\bx)\in C_\Kd$, the Cauchy problem

\begin{equation}
\pa_t\phi(\bx,t)=\langle\mu(\bx),\Lop(\bx)(\be_\mu\,\phi(\bx,t))\rangle+G(\bx,t),~~~\phi(\bx,0)\in C^{r,s},
\label{eqC}
\end{equation}

with $G\in C^{r,s}$, admits a solution which is $C^{r,s}$ w.r.t. $\bx$ and $t\in [0,T^*_0]\subset [0,T]$ for any smooth initial data. 
\end{itemize}

\brem
In applications we have that the operators 

\[\langle\be_\mu,\Lop^*(\bx)(\mu(\bx)\cdot)\rangle\mbox{ and }\langle\mu(\bx),\Lop(\bx)(\be_\mu\, \cdot)\rangle\]

 \noi are either parabolic or first order. In the parabolic case note that there is a general result (see \cite{pavliotis}), which guarantees  that if the differential operator has $C^\infty$ coefficients and the initial condition is also $C^\infty$ then the solution is 
 $C^{1,2}([0,T]\times\R^N)\cap C^{\infty}([0,T)\times\R^N)$ for some $T > 0$.
\erem

\brem
In condions (B) and (C) respectively the intervals $[0,T_0]$ and  $[0,T^*_0]$  are the maximal time intervals 
where each solution exists. Note that since solution of (B) implies (C), if (B) holds true then $T_0=T_0^*$.
\erem

\subsection{Main results}

We now state and prove the main results of the adiabatic theory for IFSS. Theorem  \ref{theoAB} is based on an elaboration of a respective proof presented in \cite{pavliotis} and clarifies the construction presented in \cite{kepler-elston}.

\bth
\label{theoAB}
For fixed $\mu\in C_\Kd$ and under assumptions (A) and (B),  equation (\ref{main-ME}) admits an asymptotic solution in each set of concentrated measures $\I_\mu$ (see definition \ref{Def:I_mu}). 
\eth

\bpf
We start with a few remarks. In the appendix we describe the geometry associated to 
the Markov chain and in particular to the kernel of $\Kd^T$ (see section \ref{geometry}).
Note any initial data in $\I_\mu$ evolve asymptotically to $\mu$. Next  we like to solve the equation

\begin{equation}
\frac{\pa \rho(\bx,t)}{\pa t}=\Lop(\rho(\bx,t))+\frac{1}{\eps}\,\Kd(\bx)\,\rho(\bx,t)
\label{eqA1}
\end{equation}

\noi by using an asymptotic expansion and conditions (A), (B). Fix $\mu \in C_\Kd$ and take an initial condition in $\I_\mu$. Let $m^*$ be an integer to be determined later. We take the expansion

\[\rho_\eps(\bx,t)=\sum_{m=0}^{m^*}\eps^m\,\rho^{(m)}(\bx,t).\]

\noi To determine $\rho^{(m)}(\bx,t)$ we substitute the expansion for $\rho_\eps$ into equation (\ref{eqA1}) and  collect the different orders in $\eps$. We obtain a hierarchy of equations

\[\begin{array}{llll}
 \displaystyle O(1/\eps):\quad \Kd^T(\bx)\,\rho^{(0)}(\bx,t)=0\\[4mm]
 \displaystyle O(1):\quad \frac{\pa \rho^{(0)}(\bx,t)}{\pa t}-\Lop^*(\rho^{(0)}(\bx,t))=\Kd^T(\bx)\rho^{(1)}(\bx,t)\\[4mm]
\displaystyle O(\eps):\quad \frac{\pa \rho^{(1)}(\bx,t)}{\pa t}-\Lop^*(\rho^{(1)}(\bx,t))=\Kd^T(\bx)\rho^{(2)}(\bx,t)\\
\vdots\\
\displaystyle O(\eps^{m^*}):\quad \frac{\pa \rho^{(m^*)}(\bx,t)}{\pa t}-\Lop^*(\rho^{(m^*)}(\bx,t))=\Kd^T(\bx)\rho^{(m^*+1)}(\bx,t).
 \end{array}\]

\noi Note that in the construction of the probability density $\rho_\eps$ with its necessary normalisation is not yet fixed. The equation (\ref{eqA1}) is linear in $\rho_\eps$. Therefore the condition

\[\int_{\R^N} d\bx\,\tr(\rho_\eps(\bx,t))=1\]

\noi must be imposed on the final form of the expansion. Conditions (A) and (B) guarantee the possibility of solving the first two equations in above hierarchy. Indeed condition (A) implies that the equation

\[\Kd^T(\bx)\rho^{(0)}(\bx,t)=0\]

\noi admits a  solution of the form

\[\rho^{(0)}(\bx,t)=f^{(0)}(\bx,t)\,\mu(\bx).\]

\noi The second equation becomes

\begin{equation}
\frac{\pa \mu(\bx)\,f^{(0)}(\bx,t)}{\pa t} -\Lop^*( \mu(\bx)\,f^{(0)}(\bx,t))=\Kd^T(\bx)\rho^{(1)}(\bx,t).
\label{eqA2}
\end{equation}

\noi By the Fredohlm alternative theorem (see \cite{pavliotis}) we have that a necessary condition for solving (\ref{eqA2}) is that the l.h.s. is orthogonal to the kernel of $\Kd(\bx)$ (see \cite{pavliotis}). Condition (A) implies  
that $\be_\mu$ satisfies

 \[ \Kd(\bx)\,\be_\mu=0.\]

\noi Therefore

\[\left\langle\be_\mu,\frac{\pa \mu(\bx)\,f^{(0)}(\bx,t)}{\pa t}-\Lop^*( \mu(\bx)\,f^{(0)}(\bx,t)) \right\rangle=\langle\be_\mu,\Kd^T(\bx)\rho^{(1)}(\bx,t)\rangle=\langle \Kd(\bx)\be_\mu,\rho^{(1)}(\bx,t)\rangle=0.\]

\noi To determine a solution one needs to solve

\[\frac{\pa}{\pa t}(\langle\be_\mu,\mu(\bx)\rangle f^{(0)}(\bx,t))-\langle\be_\mu,\Lop^*(\bx)(\mu(\bx)\,f^{(0)}(\bx,t))\rangle=0,\]

\noi which is equal to equation (\ref{eqB}) upon noting the condition

\[\langle\be_\mu,\mu(\bx)\rangle=1.\]

\noi To proceed further we compute $\rho^{(1)}(\bx,t)$. Note that any $\rho$ as a vector in $\R^g$ can be decomposed  by projection $\bPi_\mu$ (see section \ref{geometry}). Let $\rho^{(n)}(\bx,t)$ be the $n$th term of the expansion

\[\rho^{(n)}(\bx,t)=\bPi_\mu(\rho^{(n)}(\bx,t))+(\bI_\mu-\bPi_\mu)(\rho^{(n)}(\bx,t))=\xi^{(n)}(\bx,t)+f^{(n)}(\bx,t)\,\mu(\bx).\]

\noi Using that $\bPi_\mu=(\Kd^T_\mu)^D\,\bI_\mu\,\Kd^T(\bx)$, we have

\[\bPi_\mu(\rho^{(1)}(\bx,t))=(\Kd^T_\mu)^D\bI_\mu\left[\frac{\pa \rho^{(0)}(\bx,t)}{\pa t}-\Lop^*(\rho^{(0)}(\bx,t))\right].\]

\noi Since $\rho^{(0)}(\bx,t)=f^{(0)}(\bx,t)\,\mu(\bx)$ and $(\Kd^T_\mu)^D\mu(\bx)=0$ we get

\[\xi^{(1)}(\bx,t)=\bPi_\mu(\rho^{(1)}(\bx,t))=-(\Kd^T_\mu)^D\bI_\mu\,\Lop(\mu(\bx) f^{(0)}(\bx,t)).\]

\noi To construct $\rho_1(\bx,t)$ we also need that

\[(\bI_\mu-\bPi_\mu)(\rho^{(1)}(\bx,t))=\langle\be,\rho^{(1)}(\bx,t)\rangle=\mu(\bx)\,f^{(1)}(\bx,t).\]

\noi Like  for $\rho^{(0)}$ we obtain

\[\frac{\pa \rho^{(1)}(\bx,t)}{\pa t}-\Lop^*(\rho^{(1)}(\bx,t))=\Kd^T(\bx)\rho^{(2)}(\bx,t),\]

 \noi which can be projected on $\be_\mu$ leading to

\[\frac{\pa f^{(1)}(\bx,t)}{\pa t}-\langle\be_\mu,\Lop^*(\rho^{(1)}(\bx,t))\rangle=0.\]

\noi Using the fact that

\[\rho^{(1)}(\bx,t)=\xi^{(1)}(\bx,t)+\mu(\bx)\,f^{(1)}(\bx,t)=-(\Kd_\mu^T)^D\bI_\mu\,\Lop^*(\mu(\bx) f^{(0)}(\bx,t))+\mu(\bx)\,f^{(0)}(\bx,t), \]

\noi the equation for $f^{(1)}$ becomes

\[\frac{\pa f^{(1)}(\bx,t)}{\pa t}-\langle\be_\mu,\Lop^*(\mu(\bx)\,f^{(1)}(\bx,t))+\Lop^*((\Kd^T_\mu)^D\bI_\mu\,\Lop^*(\mu(\bx) f^{(0)}(\bx,t)))\rangle=0.\]

\noi The argument can be iterated. Therefore the term $\rho^{(n)}(\bx,t)$ is determined by computing its projections

\[\begin{array}{cc}
\xi^{(n)}(\bx,t):&~~~\bPi_\mu(\rho^{(n)}(\bx,t))=\xi^{(n)}(\bx,t)\\[3mm]
f^{(n)}(\bx,t):&~~~(\bI_\mu-\bPi_\mu)(\rho^{(n)}(\bx,t))=\mu(\bx)\,f^{(n)}(\bx,t).
\end{array}\]

\noi Let us assume we know $\rho^{(k)}$ from $k=0$ up to $n-1$. Then $\xi^{(n)}$ is obtained by projecting the equation

\[\frac{\pa \rho^{(n-1)}(\bx,t)}{\pa t}-\Lop^*(\rho^{(n-1)}(\bx,t))=\Kd^T(\bx)\rho^{(n)}(\bx,t),\]

\noi namely

\begin{equation}
\xi^{(n)}(\bx,t)=\bPi_\mu(\rho^{(n)}(\bx,t))=(\Kd^T_\mu)^D\bI_\mu\left[\frac{\pa \rho^{(n-1)}(\bx,t)}{\pa t}-\Lop^*(\rho^{(n-1)}(\bx,t))\right].
\label{eqxin}
\end{equation}

\noi The term $f^{(n)}$ is determined by projecting the equation

\[\frac{\pa \rho^{(n)}(\bx,t)}{\pa t}-\Lop^*(\rho^{(n)}(\bx,t))=\Kd^T(\bx)\rho^{(n+1)}(\bx,t),\]

\noi and therefore $f^{(n)}$ solves

\begin{equation}
\frac{\pa f^{(n)}(\bx,t)}{\pa t}-\langle\be_\mu,\Lop^*(\mu(\bx)\,f^{(n)}(\bx,t))+\Lop^*(\xi^{(n)}(\bx,t))\rangle=0.
\label{eqfn}
\end{equation}

\noi This concludes the construction of the expansion.\\[3mm]

\noi We now prove that the expansion of $\rho$ converges asymptotically. 
We extend a similar argument presented in \cite{pavliotis}. We show that condition (B) allows us to evaluate the regularity of the asymptotic expansion. Recall that for $n=0$

\begin{equation}
\left\{\begin{array}{ll}
\displaystyle\xi^{(0)}(\bx,t)=0\\[4mm]
\displaystyle\frac{\pa f^{(0)}(\bx,t)}{\pa t}=\langle\be_\mu,\Lop^*(\mu(\bx)\,f^{(0)}(\bx,t))\rangle
\end{array}
\right.
\label{step0}
\end{equation}

\noi holds. Also  for $n\geq 1$ we have

\begin{equation}
\left\{\begin{array}{ll}
\displaystyle\xi^{(n)}(\bx,t)=(\Kd^T_\mu)^D\bI_\mu\left[\frac{\pa \xi^{(n-1)}(\bx,t)}{\pa t}-\Lop^*(\xi^{(n-1)}(\bx,t)+\mu(\bx)\,f^{(n-1)}(\bx,t))\right]\\[4mm]
\displaystyle\frac{\pa f^{(n)}(\bx,t)}{\pa t}=\langle\be_\mu,\Lop^*(\mu(\bx)\,f^{(n)}(\bx,t))\rangle+\langle\be_\mu,\Lop^*(\xi^{(n)}(\bx,t))\rangle.
\end{array}
\right.
\label{stepn}
\end{equation}

\noi Observe that $f^{(n-1)},\xi^{(n-1)}\in C^{r,s}$, therefore equations (\ref{stepn}) and 
condition (B) imply that $f^{(n)}, \xi^{(n)}\in C^{r-2,s-1}$. The solution of (\ref{stepn}) defines the  map

\[\Psi:C^{r,s}\mapsto C^{r-2,s-1}\]

\noi as follows:

\[\Psi(\xi^{(n-1)}(\bx,t)+\mu(\bx)\,f^{(n-1)}(\bx,t))=\xi^{(n)}(\bx,t)+\mu(\bx)\,f^{(n)}(\bx,t).\]

\noi Now for $n=0$ condition (B) implies  (\ref{step0}) has a solution. Any initial condition in $C^{r,s}$  yields $f^{(0)}(\bx,t)\in C^{r,s}$. Using the map $\Psi$ we can write

\[\rho^{(n)}(\bx,t)=\Psi^n(\mu(\bx)\,f^{(0)}(\bx,t),\]

\noi with

\[\rho^{(n)}(\bx,t)\in C^{r-2n,s-n}\mbox{ for $0\leq n\leq m^*$}.\]

\noi Let us now fix the order of the asymptotic expansion to be

\[m^*=\min\left\{\frac{r-2}{2},s+1\right\}, \]

\noi so that $\rho^{(m^*)}\in C^{2,1}$. Write $\rho_\eps$ as

\[\rho_\eps(\bx,t)=\sum_{k=0}^{m^*}\eps^k\rho^{(k)}(\bx,t)+R(\bx,t),\]

\noi where $R(\bx,t)$ is the error term. Using the equation for $\rho_\eps$ we derive an equation for this error term:

\[\sum_{k=0}^{m^*}\eps^k\frac{\pa\rho^{(k)}(\bx,t)}{\pa t}+\frac{\pa R(\bx,t)}{\pa t}=\left(\Lop^*+\frac{1}{\eps}\,\Kd^T(\bx)\right)\left(\sum_{k=0}^{m^*}\eps^k\rho^{(k)}(\bx,t)+R(\bx,t)\right).\]

\noi Now using the equations for $\rho^{(k)}$ for $1<k<{m^*}$, we obtain

\[\frac{\pa R(\bx,t)}{\pa t}=\Lop^*_\eps(R(\bx,t))+\eps^{m^*}\left(
\Lop(\rho^{({m^*})}(\bx,t))-\frac{\pa \rho^{({m^*})}(\bx,t)}{\pa t}\right),\]

\noi where

\[\Lop^*_\eps\doteq \Lop^*+\frac{1}{\eps}\,\Kd^T(\bx).\]

\noi The operator $\Lop^*_\eps$ is a generator of a contraction semigroup for $t\in [0,T_0)$. Therefore we can use its exponential to compute $R(\bx,t)$, with

\[\begin{array}{ll}
\displaystyle 
R(\bx,t)=\exp{(t\,\Lop^*_\eps)}\,R(\bx,0)+\\[4mm]
\displaystyle +\eps^{m^*}\int_0^tds\exp{((t-s)\,\Lop^*_\eps)}\,\left(
\Lop^*(\rho^{({m^*})}(\bx,s))-\frac{\pa \rho^{({m^*})}(\bx,s)}{\pa s}\right).
\end{array}\]

\noi Using the semigroup property  the norm of  $\exp(t\Lop^*)$ can be bound by

\[\sup_{\bx\in\Omega,t\in [0,T_0]}\|\exp(t\,\Lop_\eps)\|=1.\]

\noi This implies the estimate

\[\sup_{\bx,t\in [0,T_0]}\|R(\bx,t)\|\leq\sup_\bx\|R(\bx,0)\|+\eps^{m^*}\int_0^{T_0}ds\,\sup_{\bx,s}\left\|\Lop^*(\rho^{({m^*})}(\bx,s))-\frac{\pa \rho^{({m^*})}(\bx,s)}{\pa s}\right\|.\]

\noi Since ${m^*}=(r-2)/2$ we have that

\[\Lop^*(\rho^{({m^*})}(\bx,s))-\frac{\pa \rho^{({m^*})}(\bx,s)}{\pa s}\in C^{0,s'},\]

\noi with $s'\geq 0$. Therefore there exists $C_1(\Omega,T_0)>0$ such that

\[\sup_{\bx\in\Omega,t\in [0,T_0]}\|R(\bx,t)\|\leq\sup_\bx\|R(\bx,0)\|+\eps^{m^*}\,T_0\,C_1(\Omega,T_0).\]

\noi Taking an initial condition satisfying

\[\sup_{\bx\in\Omega,t\in[0,T_0]}\|R(\bx,0)\|=C_2(\Omega,T_0)\,\eps^{m^*},\]

\noi the final estimation on the error is given by

\[\sup_{\bx\in\Omega,t\in [0,T_0]}\|R(\bx,t)\|\leq\eps^{m^*}(C_2(\Omega,T_0)+T_0\,C_1(\Omega,T_0)).\]

\noi Thus we get the estimate

\[\sup_{\bx\in\Omega,t\in [0,T_0]}\|\rho(\bx,t)-\rho_\eps^{({m^*}-1)}(\bx,t)\|\leq \eps^{m^*}(C_2(\Omega,T_0)+T_0\,C_1(\Omega,T_0)).\]

\noi This concludes the proof.
 \epf

\noi In case condition (B) does not hold the operator $\langle\be_\mu,\Lop^*\mu(\bx)\cdot\rangle$  does not yield a probability density  which is sufficiently smooth. In such circumstances one can look at a weaker hypothesis such as condition (C). 

\bth
\label{theoAC}

For fixed $\mu\in C_\Kd$ and under assumptions (A), (C),  equation (\ref{main-new-ME}) admits an asymptotic solution in the set of concentrated functions $\Y_\mu$. This solution gives rise to   a solution for the Kolmogorov equation and therefore to a weak solution for (\ref{main-ME}).
\eth
\bpf

The aim is to construct an asymptotic solution for

\begin{equation}
\frac{\pa u(\bx,t)}{\pa t}=\Lop(\bx)( u(\bx,t))+\frac{1}{\eps}\,\Kd(\bx)\,u(\bx,t).
\label{dual-main-ME2}
\end{equation}

\noi In the following we only outline the proof of  theorem \ref{theoAC}. The method used is close to the proof of theorem \ref{theoAB}. In summary we have the following steps:

\begin{enumerate}
\item Fix $\mu\in C_\Kd$ and consider initial conditions in $\Y_\mu$.
\item Consider an expansion of the form: $u_\eps(\bx,t)=\sum_{n=0}^{m^*}\eps^n\,u^{(n)}(\bx,t)$.
\item Construct the equation at each order $k$.
\item Decompose each $u^{(n)}(\bx,t)$ using the projection $\pi_\mu$

\[ u^{(n)}(\bx,t)=\eta^{(n)}(\bx,t)+\be_\mu\,\phi^{(n)}(\bx,t), \]

where

\[\eta^{(n)}(\bx,t)=\pi_\mu(u^{(n)}(\bx,t)),~~~\phi^{(n)}(\bx,t)=\langle\be_\mu,u^{(n)}(\bx,t)\rangle.\]

\item Construct the hierarchy of equations. For $n=0$ we have 

\begin{equation}
\left\{\begin{array}{ll}
\displaystyle\eta^{(0)}(\bx,t)=0\\[4mm]
\displaystyle\frac{\pa \phi^{(0)}(\bx,t)}{\pa t}=\langle\mu(\bx),\Lop(\be_\mu\,\phi^{(0)}(\bx,t))\rangle.
\end{array}
\right.
\label{dual-step0}
\end{equation}

\noi  Then  for $n\geq 1$ we get 

\begin{equation}
\left\{\begin{array}{ll}
\displaystyle\eta^{(n)}(\bx,t)=\Kd^D_\mu\bI_\mu\left[\frac{\pa \eta^{(n-1)}(\bx,t)}{\pa t}-\Lop(\eta^{(n-1)}(\bx,t)+\be\,\phi^{(n-1)}(\bx,t))\right]\\[4mm]
\displaystyle\frac{\pa \phi^{(n)}(\bx,t)}{\pa t}=\langle\mu(\bx),\Lop(\be_\mu\,\phi^{(n)}(\bx,t))\rangle+\langle\mu(\bx),\Lop(\eta^{(n)}(\bx,t))\rangle.
\end{array}
\right.
\label{dual-stepn}
\end{equation}

\item The evaluation of the remainder of the asymptotic series is then carried out in the same way as in theorem \ref{theoAB}.

\end{enumerate}
 \epf

\brem
It is worth to mention that in systems where $\Ld_E^*$ is not identically zero and therefore 
$\Ld^*$ is not diagonal the higher order corrections play a crucial role. In fact there exist 
systems with different $\Ld^*$ operator but same average dynamics. For such systems 
it is necessary to study also the higher order terms in the $\eps$-expansion. This 
 class of system will be investigated in a forthcoming paper.
\erem

\subsubsection*{Illustrative example: adiabatic theory and average dynamics}

We illustrate the theory by looking finally at the example from enzyme kinetics following the introduction of an IFSS. We have now obtained two different macroscopic limits due to the nature of the IFSS, in sequential order, first the continuum limit, then the adiabatic limit. The result is

\begin{equation}
\label{final_eq}
\left\{\begin{array}{ll}
\displaystyle\xi^{(n)}(\bx,t)=(\Kd^T_\mu)^D\bI_\mu\left[\frac{\pa \xi^{(n-1)}(\bx,t)}{\pa t}-\Lop^*(\xi^{(n-1)}(\bx,t)+\mu(\bx)\,f^{(n-1)}(\bx,t))\right], \\[4mm]
\displaystyle\frac{\pa f^{(n)}(\bx,t)}{\pa t}=\langle\be_\mu,\Lop^*(\mu(\bx)\,f^{(n)}(\bx,t))\rangle+\langle\be_\mu,\Lop^*(\xi^{(n)}(\bx,t))\rangle,
\end{array}
\right.
\end{equation}

where the invariant measure is 

\[\mu=\left(\begin{array}{c}
\displaystyle\frac{k^-}{k^-+a\,k^+}\\[3mm]
\displaystyle\frac{a\,k^+}{k^-+a\,k^+}\\
\end{array}
\right).\]

The matrix $\I_\mu$ id the identity in $\R^2$ and 

\[f^{(n)}(x,t)=\rho^{(n)}_0(x,t)+\rho^{(n)}_1(x,t)\mbox{ and }
\xi^{(n)}(x,t)=\left(\begin{array}{c}\xi^{(n)}_0(x,t)\\[2mm] \xi^{(n)}_1(x,t)\end{array}\right).\]

Furthermore $\Lop^*$ becomes

\[
\Lop^*= \, \left( \begin {array}{cc}
 \Dop(\nu\,x\,\cdot)  &0
 \\\noalign{\medskip}0 &  \Dop((\nu\,x-v)\,\cdot)
 \end {array} \right),\]
 
and the infnitesimal generator has the form

\[\Kd^T\, =  \left( \begin {array}{cc} 
-a\,k^+ & k^-
\\\noalign{\medskip}a\,k^+& -k^-
\end {array} \right).
\]

Finally the Drazin inverse is

\[(\Kd^T)^D= \, \frac{1}{(a\,k^++k^-)^2}\left( \begin {array}{cc} -a\,k^+& k^-\\\noalign{\medskip} a\,k^+& -k^- \end {array} \right) 
\]

\brem
The solution of (\ref{final_eq}) produces the expansion of the 
probability distribution $P(t,x,a)$. It is useful to observe that the first two terms of the expansions of $\xi$ and $f$ are given by

\[\left\{\begin{array}{lll|}
\xi^{(0)}=0, \\[4mm]
\displaystyle\frac{\pa f^{(0)}(\bx,t)}{\pa t}=\langle\be_\mu,\Lop^*(\mu(\bx)\,f^{(0)}(\bx,t))\rangle, \\[4mm]
\displaystyle\xi^{(1)}(\bx,t)=(\Kd^T_\mu)^D\bI_\mu\left[\mu(\bx)\,f^{(0)}(\bx,t))\right], \\[4mm]
\displaystyle\frac{\pa f^{(1)}(\bx,t)}{\pa t}=\langle\be_\mu,\Lop^*(\mu(\bx)\,f^{(1)}(\bx,t))\rangle+\langle\be_\mu,\Lop^*(\xi^{(1)}(\bx,t))\rangle,
\end{array}\right.\]

generate the a diffusion process  whose diffusion coefficient depend on $\eps$ and $\de$ and therefore the time evolution of the concentration $x$ will be dictated by a 
stochastic differential equation. The construction of this approximation will the subject of a forthcoming paper and is not further considered here.
\erem

The above example will be used to derive the Michaelis-Menten and Hill type kinetics known from enzyme kinetics (but also often used in genetics) as a deterministic limit of the probability distribution $P$.


\section{Appendix}
In this appendix we collect the main property of the geometrical 
property associated to the Markov chain generator $\Kd$.
\subsection{Geometry of the Markov chain}
\label{geometry}

The adiabatic approximation can be carried out by taking advantage of the geometrical structure
associated to the Markov chain, i.e. the occurence of multiple stationary measures. This will be highly relevant in applications where different parts of the Markov chain will be associated to different distinct molecular machines which will be able to exist in different modes of operation. Such a structure will be preserved by the continuum approximation leading to the most important tool to construct the adiabatic approximation of the FPE. For the construction the following definition is of importance:

\bdefi[Drazin inverse]
Let $A:\R^g\mapsto \R^g$ be a linear map with $\ker(A)\neq\emptyset$. 
 The Drazin inverse $A^D$ of $A$ is a linear map defined as
 
 \[A^D=U_A\,G_A\,U^{-1}_A,\]
 
 where
 
 \begin{enumerate}
 
\item[(i)] $G_A$ is a diagonal matrix with:
\[\begin{array}{ll}
(G_A)_{ii}=a_i\mbox{ if $a_i$ is a non-zero eigenvalue of $A$, and}\\
(G_A)_{ii}=0\mbox{ for a $0$ eigenvalue of $A$.}
\end{array}\]

\item[(ii)]  $U_A$ is the matrix whose columns are the eigenvectors of $A$. 
\end{enumerate}
\edefi

The Drazin inverse satisfies the following proposition:

\bprop
\label{drazin0}
If $v\in\ker(A)$ then $v\in\ker(A^D)$.
\eprop

\bpf
Indeed since $v$ is a column on $U_A$ we have that $U^{-1}_A\,v$ is a vector 
with all zero entries but one corresponding to $v$ in $U_A$. Therefore the 
 definition of $G_A$ implies $G_AU^{-1}_A\,v=0$. 
  \epf

The matrix $\Kd(\bx)$ is an infinitesimal generator of a finite Markov chain for every $\bx\in\R^N$, whose transpose is $\Kd^T(\bx)$.  Both  $\Kd(\bx)$ and   $\Kd^T(\bx)$ are linear operators acting on $(\R^g,\langle.,.\rangle)$. The geometric structure we are interested in is based on stationary measures:

\bdefi[Stationary measures]
\label{sm}
\[M_\Kd\doteq\left\{\mu(\bx):\Kd^T(\bx)\mu(\bx)=0,~~\sum_{i=1}^g\mu_i(\bx)=1\right\}\]
\edefi

 \noindent We make the following assumption:

\begin{itemize}
\item[($\star$)] $\dim(\ker(\Kd^T(\bx))<g$ uniformly in $\bx$. 
\end{itemize}

A trivial consequence of the definition \ref{sm} and assuption ($\star$) is:

\bprop
$M_\Kd$ is a linear subspace of $\R^g$ and $m_\Kd=\dim(M_\Kd)=\ker(\Kd^T(\bx))$. Let $\{\theta_i\}_1^g$ be a sequence of real numbers such that $\sum_{m=1}^{m_\Kd}\theta_m=1$. Then the vector

\[\mu=\sum_{m=1}^{m_\Kd}\theta_m\,\mu^{(m)}\in M_\Kd, \]
where $\mu^{(m)}\in M_\Kd$.
 \eprop

 \noindent This motivates the next definition considering convex combinations of stationary measures:

\bdefi[Convex combinations] We denote by

\[C_\Kd=\left\{\mu\in M_\Kd:\mu=\sum_{m=1}^{m_\Kd}\theta_m\,\mu^{(m)}\mbox{ with }
\sum_{m=1}^{m_\Kd}\theta_m=1,~~\theta_m\in\R_+\right\}.\]

the set of convex combinations of stationary measures if an IFSS.
\edefi

 \noindent A normalisation of the combination $\mu$ can be written as
 
\[\langle \be_\mu,\mu(\bx)\rangle=\tr(\mu(\bn))=\sum_{k=1}^g\mu_k(\bx)=1.\]

\noindent In this context it is useful to make an additional definition. First
let us introduce ${\be}_\mu^T \in \R^g$ is given by

\[(\be_\mu)_i=\left\{\begin{array}{ll}
0\mbox{ if } \mu_i=0\\
1\mbox{ if } \mu_i\neq 0,
\end{array}\right.\]

then we define 
\bdefi[Concentrated measures]
\label{Def:I_mu}
Let $\mu\in C_\Kd$. Let

\[\I_\mu\doteq\left\{ \rho : \sum_\bn\tr(\rho(\bx))=1\mbox{ and }\sum_\bn\langle\be_\mu,\rho(\bx)\rangle=1 \right\}.\]

\noi We call  $\I_\mu$ the set of concentrated measures.
\edefi

\brem
Note that the set $\I_\mu$ contains all probability distributions which have the same support as the chosen convex combination of stationary measures $\mu$.
\erem

\noi Now the vector of probabilities can be decomposed in the following way:

\bprop
\label{prop:Pi}
Given $\mu\in C_\Kd$, let $P\in \R^g$. Then $P$ can be decomposed into

\[\rho(\bx)=\xi(\bx)+ f(\bx)\,\mu(\bx), \]

 where 
 
 \[\xi(\bx)={\bPi_\mu}(\rho(\bx)),~~~ f(\bx,t)=\,{\be}_\mu^T \rho(\bn).\] 
 
 \eprop
 
 The function $f(\bx)$ is called \emph{marginal} distribution.
 
\bpf

 Let us define the operator
 
\[{\bPi}_\mu\doteq{\bI_\mu}-\mu(\bn)\,{\bf 1}_\mu^T,\]

where $\bI_\mu$ is a diagonal matrix such that $(\bI_\mu)_{\sig\sig'}=1$ if and only if 
$\mu_\sig\neq 0$ otherwise $(\bI_\mu)_{\sig\sig'}=0$. One can easily verify that

\[\bPi_\mu^2=\bPi_\mu .\]

From this relation the decomposition of $\rho(\bx)$ follows.
 \epf 

\noi The matrix $\Kd^T(\bx)$ cannot be inverted because $\ker(\Kd^T(\bx))\neq\emptyset$. 
Here we need to use the Drazin inverse. The following result holds true (see \cite{rothblum}):

\bprop
\label{drazin}
 There exists $(\Kd^T_\mu)^D$ such that 
 
 \begin{equation}
 \label{drazin-rel}
 \Kd^T(\bx)\bPi_\mu=\Kd^T(\bx)\,\bI_\mu=\bI_\mu\,\Kd^T(\bx),~~(\Kd^T_\mu)^D\,\Kd^T(\bx)={\bPi_\mu}
 \end{equation}
  \eprop
 
 \bpf
From proposition (\ref{drazin0}) follows

\[(\Kd^T_\mu)^D\,\mu(\bx)=0.\]

This proves the first relation of (\ref{drazin-rel}). For the second relation the reader is refered to \cite{rothblum}.
 \epf

\noi As the  matrix $\Kd(\bx)$ is the transpose of $\Kd^T(\bx)$ we will show that $K(\bx)$ provides a  splitting of maps from $\R^N$ to $\R^g$. More generally we shall now 
describe how to decompose any map $\Phi:\R^N\mapsto \R^g$.  This decomposition will be useful to study the weak form of the FPE. In order to formulate the decomposition we first observe the following simple implication of assumption $(\star)$:
 
\bprop 
\label{kerK}
$\ker(\Kd(\bn))$ is generated by $\{\be_\mu\}_{\mu\in M_\Kd}$.
\eprop



 \noindent Using proposition \ref{kerK} one can show that

\bprop
\label{prop:pi}
Every $\Phi:\R^N\mapsto \R^g$ can be decomposed into

\[\Phi(\bx)=\eta(\bx)+ \phi(\bx)\,\be_\mu, \]

where 

 \[\eta(\bn)={\bpi_\mu}(\Phi(\bx)),~~~ \phi(\bx)=\,{\bf 1_\mu}^T\Phi(\bx).\] 
\eprop
\bpf

Let us define:

\[\bpi_\mu\doteq\bI-\be_\mu\,\be^T\]

Note that 

\[\pi_\mu^2=\pi_\mu\]
These relations imply that the decomposition holds true.
 \epf

 \bprop
 There exits  $\Kd^D_\mu$ such that 
 
 \begin{equation}
 \label{drazin-rel2}
 \Kd(\bn)\bpi_\mu=\Kd(\bx),~~\Kd^D_\mu\,\,\Kd(\bx)={\bpi_\mu}.
 \end{equation}
 
 \eprop
 \bpf 
 The proof proceeds as in proposition \ref{drazin}.
 \epf

\subsubsection*{Illustrative example: Invariant measure and Drazin inverse}
In our illustrative example from enzyme kinetics the MC has infinitesimal generator $\Kd_\de$. Its transpose is

\[
\Kd^T\, =  \left( \begin {array}{cc} 
-a\,k^+ & k^-
\\\noalign{\medskip}ak^+& -k^-
\end {array} \right) .
\]

The invariant measure $\mu$ that satisfies $\Kd^T\mu=0$ is

\[\mu=\left(\begin{array}{c}
\displaystyle\frac{k^-}{k^-+a\,k^+}\\[3mm]
\displaystyle\frac{a\,k^+}{k^-+a\,k^+}\\
\end{array}
\right) . \]

Now the matrices $U_{\Kd}$ and $U_{\Kd}^{-1}$ are respectively 

\[U_{\Kd}= \, \left( \begin {array}{cc} 
k^-&1\\\noalign{\medskip}a\,k^+&-1\end {array} \right) 
~~\mbox{ 
and }
~~U_{\Kd}^{-1}=\left(\begin {array}{cc}  \frac{1}{a\,k^++k^-}& \frac{1}{a\,k^++k^-}\\\noalign{\medskip}{\frac{k^-}{a\,k^++k^-}}&-\frac{a\,k^+}{a\,k^++k_\de^-}
\end {array}  \right) . \]

Now

\[
G_{\Kd}\, =\left( \begin {array}{cc} 
0 & 0
\\\noalign{\medskip}0 & -\frac{1}{a\,k^++k^-}
\end {array} \right) ,
\]

so the Drazin inverse $(\Kd^T)^D=U_{\Kd}\,G_{\Kd}\,U_{\Kd}^{-1}$ is:

\[(\Kd^T)^D= \, \frac{1}{(a\,k^++k^-)^2}\left( \begin {array}{cc} -a\,k^+& k^-\\\noalign{\medskip} a\,k^+& -k^- \end {array} \right)  .
\]



\end{document}